\documentclass[%
 reprint,
nofootinbib,
 amsmath,amssymb,
 aps,
]{revtex4-2}
\usepackage{comment}
\usepackage{graphicx}
\usepackage{dcolumn}
\usepackage{bm}
\usepackage{hyperref}
\usepackage[mathlines]{lineno}

\usepackage{SIunits}
\usepackage{physics}
\newcommand{\bea}{\begin{equation}
    \begin{aligned}}
\newcommand{\eea}{\end{aligned}\end{equation}}
\newcommand{\taures}{\tau _\text{res}}
\newcommand{\taustart}{\tau _\text{start}}

\usepackage{xcolor}

\begin{document}

\preprint{APS/123-QED}

\title{
Assessing the impact of transient orbital resonances
}
\author{Lorenzo Speri}
\email{lorenzo.speri@aei.mpg.de}

\author{Jonathan R.~Gair}%

\affiliation{%
 Max Planck Institute for Gravitational Physics (Albert Einstein Institute),\\Am M\"{u}hlenberg 1, Potsdam 14476, Germany
}%

\date{\today}

\begin{abstract}
One of the primary sources for the future space-based gravitational wave detector, the Laser Interferometer Space Antenna, are the inspirals of small compact objects into massive black holes in the centres of galaxies.
The gravitational waveforms from such Extreme Mass Ratio Inspiral (EMRI) systems will provide measurements of their parameters with unprecedented precision, but only if the waveforms are accurately modeled. 
Here we explore the impact of transient orbital resonances which occur when the ratio of radial and polar frequencies is a rational number. We introduce a new Effective Resonance Model, which is an extension of the numerical kludge EMRI waveform approximation to include the effect of resonances, and use it to explore the impact of resonances on EMRI parameter estimation. 
For one-year inspirals, we find that the few cycle dephasings induced by 3:2 resonances can lead to systematic errors in parameter estimates, that are up to several times the typical measurement precision of the parameters. The bias is greatest for 3:2 resonances that occur closer to the central black hole. By regarding them as unknown model parameters, we further show that observations will be able to constrain the size of the changes in the orbital parameters across the resonance to a relative precision of $10\%$ for a typical one-year EMRI observation with signal-to-noise ratio of 20. Such measurements can be regarded as tests of fundamental physics, by comparing the measured changes to those predicted in general relativity.
\end{abstract}

\maketitle

\section{Introduction}

Gravity is the weakest of the fundamental interactions and is particularly challenging to study in a laboratory experiment.
Fortunately, the Universe has plenty of gravitational phenomena which can be observed and used to improve our understanding.
For hundreds of years we have been studying the cosmos through the observation of electromagnetic radiation, but in the last few years it finally became possible to also listen to the Universe through the observation of gravitational waves (GWs)~\citep{theligoscientificcollaborationObservationGravitationalWaves2016}.

In fact, these ripples in the fabric of spacetime allow us to study gravity and the Universe from a totally new perspective.
Indirect evidence for gravitational wave emission was first observed in 1974 by Hulse and Taylor in a study of the orbital decay of a binary pulsar \citep{taylorNewTestGeneral1982, hulseDiscoveryPulsarBinary1975}. 
The first direct observation of a gravitational wave signal was made in September 2015 \citep{theligoscientificcollaborationObservationGravitationalWaves2016}, by the ground based laser interferometer detectors LIGO \citep{aasiAdvancedLIGO2015}.
This and subsequent detections with the LIGO and Virgo detectors~\citep{acerneseAdvancedVirgoSecondgeneration2014a} have opened up a new era of gravitational wave observations 
\citep{nitz2OGCOpenGravitationalwave2020, theligoscientificcollaborationGW190425ObservationCompact2020, nitz1OGCFirstOpen2019, ligoscientificcollaborationandvirgocollaborationGW170817ObservationGravitational2017, ligoscientificcollaborationandvirgocollaborationGW170814ThreeDetectorObservation2017, ligoscientificcollaborationandvirgocollaborationGWTC1GravitationalWaveTransient2019, abbottGW170608Observation192017, ligoscientificandvirgocollaborationGW170104Observation50SolarMass2017, ligoscientificcollaborationandvirgocollaborationGW151226ObservationGravitational2016, ligoscientificcollaborationandvirgocollaborationObservationGravitationalWaves2016, theligoscientificcollaborationGW190412ObservationBinaryBlackHole2020a}, which have permitted studies of the properties of the astrophysical population of compact objects, such as black holes and neutron stars, and have led to some of the most stringent tests of General Relativity. 

The ground based gravitational wave detectors have a frequency range from about $10$Hz to a few kHz, which permits observations of merging binaries of stellar mass compact objects at low redshift, $z < 2$. 
The European Space Agency plans a new space-based gravitational wave observatory, the Laser Interferometer Space Antenna (LISA), to open a \emph{gravitational window} on the Universe in the low-frequency range of $10^{-4} - 10^{-1}$ Hz.
LISA is expected to resolve thousands of overlapping gravitational wave signals from a wide variety of sources \citep{2017arXiv170200786A}.

The inspiral of a stellar-origin compact object (CO) into a massive black hole (MBH) in the centre of a galaxy is one of the potential sources of gravitational waves for LISA.
Observations of such Extreme Mass Ratio Inspiral (EMRI) systems have a huge scientific potential. The compact object typically completes $10^4 - 10^5$ cycles in band, during which time it is orbiting in the strong field region close to the central black hole. 
Therefore, EMRI signals encode a detailed map of the background space-time of the central MBH and offer a unique opportunity to measure the properties, evolution and environment of MBHs 
\citep{barackLISACaptureSources2004,arunMassiveBlackHole2009, barausseCanEnvironmentalEffects2014, gairConstrainingPropertiesBlack2011,gairLISAExtrememassratioInspiral2010,amaro-seoaneIntermediateExtremeMassRatio2007}, to test for deviations from General Relativity (GR)~\citep{gairTestingGeneralRelativity2013,barackUsingLISAEMRI2007} and to constrain cosmological parameters \citep{macleodPrecisionHubbleConstant2008,laghi2021gravitational}.
We expect to observe between a few and a few hundred EMRIs over the LISA mission duration \citep{gairEventRateEstimates2004,gairProbingBlackHoles2009,babakScienceSpacebasedInterferometer2017,pan2021formation}, with the large uncertainty driven by uncertainties in the rates of the various dynamical processes that can lead to the capture of compact objects by a MBH \citep{amaro-seoaneIntermediateExtremeMassRatio2007, amaro-seoaneRelativisticDynamicsExtreme2018}.

In order to realize the scientific potential of these astrophysical sources we have to correctly detect and characterize EMRI signals.
A sequence of LISA data challenges showed the feasibility of identifying individual EMRIs in data sets without other sources, and using narrow parameter priors \citep{babakMockLISAData2008a, babakAlgorithmDetectionExtreme2009, babakMockLISAData2010}.
However, any EMRI data analysis technique will rely on the existence of waveforms that are both accurate enough to faithfully identify astrophysical EMRI signals and that are quick enough to generate in the large number required to search the wide parameter space of potential signals (see for example recent work in ~\citep{Chua:2020stf,Katz:2021yft}).

Accurate modeling of EMRI waveforms can be accomplished within the framework of black-hole perturbation theory, regarding the mass ratio $\eta \sim 10^{-4} -10^{-7}$ as a small expansion parameter. 
Within this framework, the evolution of the orbit of the compact object is governed by the Kerr geodesic equations with a forcing term, called the gravitational self-force, which takes into account the finite size and mass of the body and its back reaction on the background Kerr spacetime \citep{PhysRevD.55.3457,PhysRevD.56.3381}. 
Generic EMRI orbits have three fundamental frequencies: the radial frequency $\omega_r$, the polar frequency $\omega_\theta$ and the azimuthal frequency $\omega_\phi$. 
When the radial and polar frequencies become rational multiples of each other, a transient orbital resonance occurs and the motion of the small body is restricted to a subspace within the full orbital torus~\citep{flanaganTransientResonancesInspirals2012}. The standard adiabatic approximation to compute EMRI phase space trajectories, which involves averaging over the full torus of the orbit in phase space, is then no longer valid.
Instead, additional terms in the evolution equations must be taken into account, which cause the system to exhibit a qualitatively different behavior, which has been proven to affect the detectability of EMRIs \citep{berryImportanceTransientResonances2016}.

Our work aims to investigate two questions related to EMRI parameter estimation in the presence of resonances. The first is to explore the size of parameter biases that arise when the presence of resonances is ignored while building a waveform model. The second is to explore with what precision the effect of the resonance, i.e., the amount by which the constants describing the orbit change over the resonance, can be measured from an observation. EMRI resonances have been studied in a number of previous works~\cite{vandemeentConditionsSustainedOrbital2014,vandemeentResonantlyEnhancedKicks2014,bongaTidalResonanceExtreme2019, Brink_2015,lukesgerakopoulos2021nonlinear, flanaganResonantlyEnhancedDiminished2014,ruangsriCensusTransientOrbital2014,PhysRevD.88.023002, PhysRevD.83.104024, 10.1093/ptep/pty136}, but this paper is the first attempt to explore resonant EMRI parameter estimation.

We achieve this goal using an Effective Resonance Model. This is a phenomenological model that adds transient orbital resonances into the Numerical Kludge (NK) model framework, which is an approximate waveform model that can quickly generate generic EMRI inspirals and compute their GW signals \citep{babakKludgeGravitationalWaveforms2008, gairImprovedApproximateInspirals2006}.

This work is organized as follows: in Section \ref{sec:emri_modeling} we review the 
general approach to EMRI modeling, we briefly describe the Numerical Kludge model and we present the GW data analysis tools we employ.
Subsequently, in Section~\ref{sec:emri_resonance}, we review the theoretical understanding of transient orbital resonances which will be used to construct the Effective Resonance Model in Section~\ref{subsect:eff_res_model}.
Finally, in Section~\ref{sec:results}, we show the results of our investigation using the new resonance phenomenological models. We compute the dephasings induced by 3:2 resonances over the parameter space and assess the corresponding parameter biases that would be induced by ignoring resonances, and we explore the measurability of resonance effects over a wide range of resonance strengths.

\paragraph*{Notation}
Throughout, we will use a spacelike signature $(-,+,+,+)$ for the metric and work with a system of geometrized units in which $G=c=1$. We use the Einstein summation convention for repeated indices. Greek letters will be used to indicate a sum over all spacetime indices, whereas latin letters will be used to indicate a sum over spatial only.

\section{EMRI Modeling}
\label{sec:emri_modeling}
In this section we give a general overview of EMRI waveform modeling, and highlight how this should be modified to include the effects of resonances. We will also summarise how the Numerical Kludge model is constructed. Lastly, we will review some tools from gravitational wave data analysis that we will use to assess the impact of resonances on gravitational waves from EMRIs.

\subsection{From Kerr geodesics to adiabatic evolution}

Kerr geodesics are a good approximation to the evolution of EMRIs on short time scales (for order of several orbital periods). Their properties are also a useful basis for understanding the motion of the compact object at higher orders in the mass ratio.
From the symmetries of the Kerr space-time, it is possible to identify four integrals of the geodesic equations of motion: the energy, $E$, the axial angular momentum, $L_z$, Carter's constant, $Q$ and the mass of the compact object, $\mu$.
The first two are direct consequences of the stationarity and axial symmetry of Kerr space-time.
Carter's constant is a conserved quantity associated with a tensorial Killing vector field \cite{carterGlobalStructureKerr1968} and, in the weak field or non-spinning limits, it corresponds to the squared angular momentum component parallel to the equatorial plane ($Q \approx L_x ^2 + L_y ^2$).
The mass is conserved because of the normalization of the four-momentum $ g_{\text{Kerr}} ^{\nu \mu} p_\nu p_\mu = -\mu ^2 $, i.e., because of the conservation of the Hamiltonian $H =  g_{\text{Kerr}} ^{\nu \mu} p_\nu p_\mu /2$.

The existence of four constants of motion $J_\beta = \qty(E, L_z, Q, \mu)$ allows us to reduce the geodesic equations from four second-order differential equations to four first-order equations. In Boyer-Lindquist coordinates,~$(t,r,\theta,\phi)$, the first-order equations are:
\bea
\label{eq:geodesics}
\Sigma  ^2 (r,\theta) \, \qty (\dv{r}{\tau})^2 =& 
\left[E\left(r^{2}+a^{2}M^2\right)-a M L_{z}\right]^{2} \\
& -\Delta\left[r^{2}+\left(L_{z}-a M E\right)^{2}+Q\right]
\\
\Sigma^2 (r,\theta) \, \qty (\dv{\theta}{\tau})^2 =\,& Q-\cot ^{2} \theta L_{z}^{2}-a^{2} M^2 \cos ^{2} \theta\left(1-E^{2}\right) \\
\Sigma(r,\theta) \, \dv{\phi}{\tau} =\,& 
\csc ^{2} \theta L_{z}+a M E\left(\frac{r^{2}+a^{2} M^2}{\Delta}-1\right)\\ &-\frac{a^{2} M^2 L_{z}}{\Delta} \\
\Sigma(r,\theta) \, \dv{t}{\tau} =\,&  E\left[\frac{\left(r^{2}+a^{2} M^2\right)^{2}}{\Delta}-a^{2}M^2 \sin ^{2} \theta\right] \\
&+a M L_{z}\left(1-\frac{r^{2}+a^{2}M^2}{\Delta}\right)
\eea
where $\Sigma(r,\theta) = r^2 + a^2 M^2 \cos^2 \theta$, $\Delta = r^2 -2 Mr+a^2 M^2$ and $M$ and $a$ are the Kerr black hole mass and dimensionless spin parameter, respectively, and $\tau $ is the proper time. By using the Carter-Mino time parameter $\lambda$ defined via $d \tau /d\lambda = \Sigma$ \citep{carterGlobalStructureKerr1968,PhysRevD.67.084027}, the equations in $r$ and $\theta$ decouple and the radial and polar motion can be determined if constants and initial conditions are specified.

The trajectory of EMRIs can be modelled with Kerr geodesics only on a short time scale, of the order of the orbital time scale $\sim O(1)$, compared to the inspiral time scale $\sim O(1/\eta)$.
If we want to describe the long inspiral of EMRIs, it is necessary to take into account the impact of the gravitational field of the compact object on the background Kerr space-time. During the inspiral, the compact object slowly deviates from geodesic orbits and this can be interpreted as an effective acceleration/force due to the so-called Gravitational Self-Force (GSF) acting on the compact object \cite{barackSelfforceRadiationReaction2019}.

By using the \emph{Action-Angle} formalism it is possible to rewrite the geodesic equations to include the effect of the first order GSF as \cite{barackSelfforceRadiationReaction2019,hindererTwoTimescaleAnalysis2008}:
\begin{subequations}\begin{align}
\dv{q_\alpha}{\tau} &= \omega_\alpha (J_\beta) + \eta \, g_\alpha(q_r, q_\theta, J_\beta) + \order*{ \eta ^2}\label{eq:angle_evolution}\\
\dv{J_\alpha}{\tau} &=  0 + \eta \, G_\alpha(q_r, q_\theta, J_\alpha) + \order*{ \eta ^2} \, \label{eq:constant_evolution},
\end{align}
\end{subequations}
where the evolution of the Boyer-Lindquist coordinates is described using the angle variables $q_\alpha = (q_t,q_r,q_\theta,q_\phi)$ \cite{fioraniLiouvilleArnoldNekhoroshev2003} and the corrections due to the GSF are encoded in $G$ and $g$. We note that here the action variables, $J_\alpha$, are not precisely the same as the standard constants of motion introduced earlier, $J_\beta$, but are related by a transformation. This distinction is not important for the present discussion.

At ``zeroth'' order in $\eta$ the motion is completely determined by the fundamental frequencies $\omega_\alpha$, that define the fundamental modes of the evolution of the system and reduce to the usual Keplerian frequencies in the Newtonian limit. Expressions for these quantities can be found in \cite{schmidtCelestialMechanicsKerr2002,Fujita_2009}.
The radial fundamental frequency $\omega_r$ is associated with the radial motion and it is zero for circular orbits. The polar and azimuthal fundamental frequencies, $\omega_\theta$ and $\omega_\phi$, are associated to precessional motion of the orbit.

The four constants of motion $J_\alpha$ determine the ``shape" of the orbit, whereas the phase variables give information on the (time dependent) location of the object within the orbit, and the orbit's orientation  \cite{barackSelfforceRadiationReaction2019}.
The \emph{Action-Angle} formalism not only makes the periodicities of the system obvious, but it also facilitates the decomposition of any dynamical field as an expansion in the fundamental frequencies.

The equations~(\ref{eq:angle_evolution}-\ref{eq:constant_evolution}) can be further simplified by using the \emph{two-time scale expansion} developed by Flanagan and Hinderer \cite{hindererTwoTimescaleAnalysis2008}, which consists of separating the evolution of the EMRI into two time scales, a long time scale $J/\dot{J} \sim O(1/\eta)$ associated with the evolution of the constants of motion, and a short time scale~$q/\omega \sim O(1)$ associated with the evolution of the angle variables.
By taking an average of the flux equation (\ref{eq:constant_evolution}) it is possible to simplify the evolution of the constants of motion:
\bea
\label{flux-average}
\Big \langle \dv{J}{\tau} \Big \rangle_{q_r\, q_\theta}  &=
\eta \,G_{0,0}  (J) +\eta \,
\sum_{l\neq 0 , m\neq 0} G_{l,m}  (J) \langle  e^{i (l q_r + m q_\theta)} \rangle_{q_r\, q_\theta} \\ 
&\approx \eta \,G_{0,0}  (J) .
\eea
where we have used an expansion of $G_\alpha(q_r, q_\theta, J_\alpha)$ as a Fourier series and have introduced the 2-torus average over the phase variables, defined by
$$
\langle f \rangle_{q_r\, q_\theta} = \frac{1}{(2\pi)^2 } \int _0 ^{2 \pi} \int _0 ^{2 \pi} 
f(q_r , q_\theta) \, \dd q_r \,\dd q_\theta \, .
$$
For ergodic trajectories, i.e., phase-space filling trajectories, the phase variables evolve rapidly and the exponential term averages to zero $\langle  e^{i (l q_r + m q_\theta)} \rangle_{q_r\, q_\theta} \approx 0$ \cite{hindererTwoTimescaleAnalysis2008}.

The adiabatic approximation consists of using the averaged equations for the evolution of the constants of motions. In this way, the evolution of averaged constants can be found independently of the evolution of the phases. The resulting $J_\alpha(t)$ can then be used to solve equation~(\ref{eq:angle_evolution}) for the phase variables by also dropping the oscillating terms $g$.

However, if the argument of the exponential does not rapidly oscillate, the average cannot be taken. As a consequence, the adiabatic approximation breaks down and it is necessary to take into account another secular term. This is precisely what happens when resonances occur, and we will return to this in Section~\ref{sec:emri_resonance}.\\

\subsection{Numerical Kludge Model for EMRI Waveforms}
\label{NK}

The Numerical Kludge (NK)~\cite{babakKludgeGravitationalWaveforms2008} is one of a number of fast but approximate EMRI models that have been developed to substitute the computationally expensive GSF calculations when exploring EMRI data analysis. 
Other models include the Analytical Kludge (AK) \cite{barackLISACaptureSources2004}, the Augmented Analytical Kludge (AAK) \cite{chuaFastFiducialAugmented2017, Chua:2020stf,Katz:2021yft} and the Near Identity Transform (NIT) \cite{vandemeentFastSelfforcedInspirals2018}.

The Numerical Kludge (NK) generates a waveform for a one-year generic EMRI inspiral, sampled with a $10$s time step, in about one minute, and its implementation is well suited to include resonances. 
Here we briefly summarise how the NK is constructed. The model uses a Keplerian parameterization of the orbit to simplify the treatment of turning points in the $r$ and $\theta$ motion \cite{schmidtCelestialMechanicsKerr2002,PhysRevD.69.044015}
\bea
r &= \frac{p\, M}{1+e\cos \psi} \qquad \qquad r_p \leq r \leq r_a \\
\cos \theta &= \cos \theta_- \cos \chi \qquad \qquad \theta_- \leq \theta \leq \pi - \theta_- \, .
\eea
Here $r_p$ and $r_a$ denote the turning points of the orbits at periapsis and apoapsis, and we have introduced the eccentricity, $e$, and semi-latus rectum, $p$, which are defined in terms of the turning points through $r_p=p/(1+e)$, $r_a = p/(1-e)$. We denote the relativistic anomaly by $\psi$ and the ``polar phase'' by $\chi$.
The latter two are $2 \pi$ periodic but they are not canonical phase variables.
If we replace the Carter constant by an inclination angle
\footnote{Prograde orbits have $0\leq \iota < \pi/2 $, retrograde orbits $\pi /2 < \iota < \pi $, equatorial $\iota = 0, \, \theta_- = \pi/2 $ and polar orbits $ \iota = \pi/2,\, \theta_- = 0$.}
 defined by $\tan \iota = \sqrt{Q}/L_z$, the  motion can be described in terms of $(p, e, \iota)$ instead of $(E, L_z, Q)$ and the flux evolution becomes:
\bea
\dv{J}{t} = f_{\text{NK}} \qty(a, M, \mu, p, e, \iota ) \, .
\eea
The computation of the gravitational waveform proceeds via the following steps:
\begin{itemize}

  \item[$(i)$] The parameters defining the waveform can be grouped in the following sets:
  \begin{itemize}
    \item intrinsic parameters
    $$
    (p/M, \log \eta, \log M, e, \chi_0,  \iota, \psi _0,  a) \, ,
    $$
    where $\chi_0, \psi _0$ are the initial phases and $a$ is the dimensionless spin parameter of the central BH.
    The mass $M$ of the central MBH is the mass observed in the detector frame and it is related to the source mass $M_{s}$ through the redshift $z$ by $M = (1+z) M_{s}$.
    \item extrinsic parameters:
    $$
    (\theta_K, \phi_K, \theta_S, \phi_S, \phi_0, D_L) \, ,
    $$
    where the two sky-position angles $(\theta_S, \phi_S)$ are the co-latitude and azimuth in an ecliptic based coordinate system, the two angles $(\theta_K, \phi_K)$ define the direction of the spin of the massive black hole, with respect to the same coordinate system as the sky position\footnote{This is the direction on the sky to which the spin would point, if the black hole was transported to the Solar System Barycentre.}, $\phi_0$ is the azimuthal initial phase and $D_L [\text{Mpc}]$ is the luminosity distance.
    \item the length of the inspiral $t_\text{max}$, and the sampling interval $\Delta t$.

  \end{itemize} 

  \item[$(ii)$] The evolution of the constants of motion $J = (E, L_z, Q)$ is computed using the adiabatic approximation, i.e., replacing the first order radiative self-force $G$ with its 2-torus averaged value $f_{\text{NK}} = \langle G \rangle _{q_r \, q_\theta}$, and neglecting the conservative piece of the GSF. The two torus average is approximated with the dissipative part of the first order self-force, obtained from second order post-Newtonian formulae and fits of Teukolsky-based inspirals \cite{poundLimitationsAdiabaticApproximation2005, gairImprovedApproximateInspirals2006}. We will have to modify this flux evolution equation to take into account resonances.
  
  \item[$(iii)$] The inspiral trajectory is calculated by integrating the Kerr geodesic equations, evaluated for the instantaneous constants of the motion computed from the evolution equations and dropping all the forcing terms $g_\alpha$ in Eq.~(\ref{eq:angle_evolution}). Then, we identify the Boyer-Lindquist coordinates $(r_{BL},\theta_{BL},\phi_{BL}) \rightarrow (r_{\text{sph}}, \theta_{\text{sph}}, \phi _{\text{sph}})$ with a set of flat-space spherical polar coordinates. By doing this, the particle is effectively forced to move along a curved path in flat spacetime.
  \item[$(iv)$]
  The gravitational waveform is then calculated by applying the quadrupole formula in the TT gauge to the pseudo-flat-space particle orbit, obtaining $h_+$, $h_\times$, which can then be projected into the observer direction.
  \item[$(v)$] The low-frequency approximation to the LISA response function, described in \cite{cutlerAngularResolutionLISA1998,barackLISACaptureSources2004}, is then used to transform the waveform polarizations $h_{+,\times}$ into the two LISA response functions $h_I$ and $h_{II}$.
\end{itemize}

Despite the several approximations, the NK waveforms have been shown to be remarkably faithful, for periastron $r_p \gtrapprox 5 M$, when compared to Teukolsky-based inspirals \cite{babakKludgeGravitationalWaveforms2008}. The NK procedure works well because it takes into account the most important physics, allowing phenomenological waveforms for generic EMRIs to be quickly generated without losing too much faithfulness~\cite{babakKludgeGravitationalWaveforms2008}.

\subsection{Waveform analysis methods}

Detection and parameter estimation of gravitational wave signals employs a wide range of techniques used to identify and characterize one or more signals, $h(t)$, present in the output, $s(t)$, of a gravitational wave detector.
We assume that the output of a gravitational wave detector is composed of a signal $h(t; \vb*{\lambda} )$, dependent on the source parameters, $\vb*{\lambda}$, and instrumental noise $n(t)$:
\bea
s(t) = h(t ;\vb*{\lambda} ) + n(t). \qquad \qquad 
\eea
By assuming that the noise is weakly stationary, Gaussian and ergodic with zero mean, we can write the likelihood for the parameters $\vb*{\lambda}$ as \cite{whittleAnalysisMultipleStationary1953}:
\bea
\label{eq:likelihood}
p(s|\vb*{\lambda}) \propto \exponential \qty [-\frac{1}{2} \braket{s - h(\vb*{\lambda})}{s- h(\vb*{\lambda})} ]\, ,
\eea
where we have introduced the inner product $\braket{\cdot}{\cdot}$
$$
\braket{a (t)}{b (t)} =4 \Re \int _{0} ^\infty \frac{\tilde{a} ^* (f) \tilde{b} (f) }{S_n (f)} \, \dd f \, .
$$
The tilde indicates the continuous Fourier transform and $S_n (f)$ is the one-sided spectral density of noise in the detector.
From a practical point of view the spectral density represents our information on the detector sensitivity. In this work we adopt the LISA PSD described in \cite{robsonConstructionUseLISA2019}.

Given a waveform $h$, the optimal matched filtering signal-to-noise ratio (SNR) is defined by
$$
\text{SNR} = \rho = \sqrt{\braket{h}{h}} = \qty[ 4 \int _{0} ^\infty \frac{ \abs{\tilde{h}  (f)}^2 }{S_n (f)} \, \dd f ]^{\frac{1}{2}}\, .
$$
This is a measure of the detectability of a particular signal. The precision with which observations will be able to determine the parameters of a system can be estimated using the Fisher Information Matrix, which is 
\bea
\label{eq:fisher_matrix_def}
\Gamma _{ij} &= \mathbb{E} \qty [ 
    \pdv{l}{\lambda^i} \, \pdv{l}{\lambda^j}
]
= \braket{\partial_i h(t;\vb*{\lambda})}{\partial _j h(t;\vb*{\lambda})}
\, ,
\eea
where $l$ is the log-likelihood, $\mathbb{E}$ is the expected value, $\partial_i \equiv \partial/\partial \lambda_i$ and to obtain the second line we we have used the likelihood given in eq.~(\ref{eq:likelihood}).
Formally, the Fisher Matrix provides a lower bound on the variance of any unbiased estimator $\hat{\lambda}$ of the parameters of the signal, but it is also provides a Gaussian approximation to the shape of the likelihood, valid in the high SNR limit. The square-roots of the diagonal elements of the inverse Fisher Matrix thus provide an estimate of the precision with which the corresponding parameter can be measured in an observation, $\Delta \lambda ^i = \sqrt{\qty(\Gamma^{-1}) _{ii}}$.

If we use an approximate waveform model $h_m(\vb*{\lambda})$ to estimate the parameters $\vb*{\lambda}_0 $ of a signal actually described by a model $h_t(\vb*{\lambda})$, the recovered parameters will be affected by systematic errors. In the linear signal approximation, the shift in the peak of the likelihood due to statistical and systematic errors is given by \citep{vallisneriStealthBiasGravitationalWave2013}:
\bea
\label{eq:measurement_err}
    \delta ^i &= \underbrace{
        \qty( \Gamma ^{-1} )^{ki} \braket{\partial_k h(\vb*{\lambda}_0)}{n}
    }_{\text{statistical error} = \delta\lambda^{\text{stat} }} \\
    &+\underbrace{ 
        \qty( \Gamma ^{-1} )^{ki} \braket{\partial_k h(\vb*{\lambda}_0)}{
        h_t(\vb*{\lambda}_0) - h_m(\vb*{\lambda}_\text{BF}) }     
    }_{\text{systematic error}= \delta \lambda^{\text{bias} }} %
     \, .
\eea

The statistical error is determined by the noise realisation and scales with the SNR as $\delta\lambda^{\text{stat} }\sim 1/\rho$.
By contrast, the systematic error does not depend on the SNR and encodes the bias on the recovered parameters that arises from mis-modeling. 
If the bias from ignoring a resonance is smaller than the statistical error, then the effects of the resonance are negligible.\\

The inner product can be used to define an overlap $\mathcal{O}(a,b)$ in the usual way: 
$$
\mathcal{O} (a,b) = \frac{\braket{a}{b}}{\sqrt{\braket{a}{a}} \sqrt{\braket{b}{b}}}.
$$
This expresses how ``similar'' two signals $a(t)$ and $b(t)$ are. If two signals are identical then the overlap is 1.
We define also the Mismatch $\mathcal{M}$ as $\mathcal{M} = |1 - \mathcal{O}|$

We will often compare two kinds of waveforms: resonant waveforms which are signals produced with the NK including the Effective Resonance Model and non-resonant waveforms which are produced with the standard NK model.
We will focus on a specific subset:
$$
\vb*{\lambda} = (p/M, \log \eta, \log M, e,\iota, a, \mathcal{E}, \mathcal{L}_z, \mathcal{Q}) \, ,
$$
of the full parameter space, where the resonance coefficients $\mathcal{C} =( \mathcal{E}, \mathcal{L}_z, \mathcal{Q})$ will be defined in the next section.
These are new parameters of the waveform, which are set to zero for non-resonant waveforms, i.e., if we want to ignore resonances. Note that we are not including the phase angles or extrinsic parameters in this list. This is for computational convenience, as the accurate evaluation of a Fisher Matrix on the full EMRI parameter space is very challenging. While the phase angles will correlate with other parameters, we expect our conclusions to be minimally impacted by ignoring them. Passing through a resonance leads to a change in the orbital frequencies relative to a non-resonant trajectory, and hence a growing phase discrepancy after the resonance. This growing phase difference is likely to dominate over any small phase differences that accumulate during the resonance itself, and will not be degenerate with phase offsets encoded in $\psi_0$ and $\chi_0$.

In all the studies reported here we have set the noise realisation, $n=0$, since our focus is on the systematic biases rather than the statistical ones.
We also set to zero the initial phases and choose the extrinsic parameters to be 
$(\theta_K, \phi_K, \theta_S, \phi_S, D_L) = (\pi/8, 0, \pi/4, 0, 200 \text{ Mpc})$.
After computing a waveform with these parameters, we then rescale the distance in order to fix the signal-to-noise ratio to 20. This is thought to be roughly the detection threshold for EMRIs and hence will be a fairly typical value for observed systems. For the remainder of this study all events will have SNR of 20.
For further technical details and a report of tests used to validate our Fisher Matrix calculations, we refer the reader to the appendices \ref{app:data_analysis} and \ref{app:fish_validation}.

\section{Transient Orbital Resonances}
\label{sec:emri_resonance}

Resonances are a common phenomenon in nature in which there is a change in the evolution of some physical quantity due to coherent superposition of a forcing term.
When two frequencies of a system become commensurate, i.e., their ratio is a rational number, a resonance can occur and induce a distinctive change in the evolution. 
Flanagan and Hinderer found that transient orbital resonances occur in EMRIs when the radial frequency $\omega_r$ and the polar frequency $\omega_\theta$ become commensurate \cite{flanaganTransientResonancesInspirals2012}.
During a resonance, the phase-space trajectory is not space-filling and therefore the averaging procedure of the adiabatic approximation breaks down. This leads to a deviation of the evolution of the orbital constants from the adiabatic trajectory. 
The phase evolution also shifts away from the standard evolution leading to an overall dephasing in the emitted gravitational waves.
To study the impact of these resonances, it is necessary to build a model that takes them into account and that can be quickly used to study their impact. To this end, we review in the next sections the main features of transient orbital resonances in EMRIs and use them as a basis to construct an Effective Resonance Model. This phenomenological model for resonances is then implemented in the NK model framework and used to investigate the impact of these phenomena on parameter estimation.

\subsection{Properties}
The evolution of the constants of motion (eq.~(\ref{eq:constant_evolution})) can be decomposed in a Fourier series in the radial and polar angle variables, $q_r$ and $q_\theta$.
Using this expansion, the change in the fluxes is manifestly expressed by two terms: the secular term $G_{0,0}$ and the rapidly oscillating term.
The latter one can be averaged out using the adiabatic approximation for ergodic trajectories.
However, if the phase  $\Psi = l q_r + m q_\theta$ of the oscillatory term is stationary, the adiabatic approximation is no longer valid and additional contributions add up to the secular term $G_{0,0}$.
By expanding the phase variables in terms of the fundamental frequencies, we can understand when this occurs:
\bea
\label{resonance-expansion}
\Psi = 
&\overbrace{ l \,q_{r 0} + m \,q_{\theta 0} } ^\text{$\Psi_0$ phase at resonance}+ 
\overbrace{ \qty(l \,\omega_{r 0} + m\, \omega_{\theta 0}) }^{\text{resonance condition}} (\tau - \tau_0) \\
&+ 
\underbrace{ \frac{1}{2} \qty(l \, \dot{\omega}_{r 0} + m \, \dot{\omega}_{\theta 0})  }_{\propto \tau^{-2} _\text{res} } \, (\tau - \tau_0)^2 + \dots
\eea
The phase $\Psi$ becomes stationary in the proximity of a resonance, i.e., when the resonance condition: $l^* \,\omega_{r 0} + m^*\, \omega_{\theta 0} = 0$ is satisfied around $\tau = \tau_0$ for a given Fourier mode $l^*,m^* \in \mathbb{Z} $. Therefore, when the ratio of the radial and polar fundamental frequencies, $m^* / l^*$, is a rational number, the respective term $G_{l^*,m^*}$ and all the higher harmonic terms $s\, l^*,s\, m^*$, for $s \in \mathbb{Z}$, (i.e., integer multiples of $l^*,m^*$) are also approximately secular and the adiabatic approximation breaks down. In fact, during a resonance, the phases $q_r,\, q_\theta$ cover only a specific non-ergodic trajectory and the phase-space average of the normally oscillating term is non zero. The flux equation (\ref{flux-average}) during a resonance becomes: 
\bea
\label{eq:flux-resonant}
\Big \langle \dv{J}{\tau} \Big \rangle_{q_r\, q_\theta}  
&=
\eta \,\qty[G_{0,0}  (J) +
\sum_{s \neq 0 } G_{s \,l^*,s\, m^*}  (J) \, e^{i \,s(l^* q_r + m^* q_\theta)}   ] \\
&=\eta \,G_{0,0}  (J) + \eta \,G_{l^*,m^*} e^{i\Psi_0} e^{2 \pi \, i \qty(\frac{\tau - \tau_0 }{\taures} )^2} + \dots
\eea
The resonance ends when the phase $\Psi$ starts oscillating again, therefore, when the third term in equation (\ref{resonance-expansion}) is significantly different from zero.
Thus we can define the duration of the resonance as: 
$$
\tau_\text{res} = \sqrt{2 \frac{2 \pi}{\abs{l^* \, \dot{\omega}_{r 0} + m^* \, \dot{\omega}_{\theta 0}} }} \, .
$$
We want to stress that there is not a sharp transition between the resonance regime and adiabatic regimes but a smooth one and, therefore, the resonance duration is somehow an arbitrary definition up to overall factors.

Since the resonance condition could be satisfied for every pair of integers $(l,m)$, one might think that an EMRI is basically always on resonance.
However, only the low order resonances are significant.
In fact, if the resonance order is high it means that the number of cycles in $r$ with respect to the cycles in $\theta$ is high enough that the average over the 2-torus is effectively done.
Furthermore, the duration of the resonance scales as the inverse of the order of the resonance $|l| + |m|$ and only low order resonances are long enough to affect the evolution.

The fundamental frequencies and the constants of motion change typically on a time scale of $J/\dot{J} \sim \omega/\dot{\omega} \sim 1/\eta$. 
When an EMRI system passes through a resonance, the evolution of the constants of motion deviates from the standard adiabatic evolution.
This deviation is caused by the additional contributions to the fluxes in Eq.~(\ref{eq:flux-resonant}) and its size is of order $\Delta J \sim \dot{J} \tau_\text{res} \sim \sqrt{\eta}$, and analogously for the fundamental
frequencies $\Delta \omega \sim \sqrt{\eta}$.
Therefore, the evolution of the constants of motion receives a ``kick", which can be either positive or negative depending on the phase at resonance $\Psi _0$ and on $\,G_{l^*,m^*}$.
This deviation is bigger than any higher order corrections in the mass ratio.
The exact expression for the total change in the constants of motion after resonance $l^*$:$m^*$ can be found in \cite{berryImportanceTransientResonances2016,flanaganTransientResonancesInspirals2012}.
The jumps in the constants cause the phases to shift away from the standard adiabatic evolution. 
This leads to a cumulative dephasing of order $\Delta q = \Delta \omega \, \mathcal{T} \sim 1/\sqrt{\eta}$ by the end of the inspiral, where the inspiral timescale $\mathcal{T} \sim 1/\eta$.
These shifts of the phases can be particularly large in small mass ratio systems, which motivates investigating their effect on EMRI parameter estimation. To do so we need to build a model for resonances.\\

\subsection{Effective Resonance Model}
\label{subsect:eff_res_model}
Here we develop a phenomenological model of resonances based on the main features of transient orbital resonances discussed in the previous section.
This is done without reducing the computational efficiency of the Numerical Kludge.
We build our model with two main ingredients: a criterion to specify when a resonance starts, and a modification of the flux equation during resonance.
The Numerical Kludge evolves firstly the phase-space trajectory of the constants of motion $J =(E, L_z,Q)$ and, since the fundamental frequencies are only functions of $J$, the condition for resonance can be checked at each time step. If it is satisfied, we modify the flux evolution to include the effects of the resonance.

The 3:2 resonances are considered to have the strongest impact on the inspiral \cite{berryImportanceTransientResonances2016}, and are encountered by EMRI systems long enough before plunge that a significant phase shift can accumulate \cite{ruangsriCensusTransientOrbital2014}.
We will therefore explain how to construct the Effective Resonance Model for the case of a 3:2 transient orbital resonance, but the extension to other resonances is straightforward. All frequencies and times will refer to coordinate time, unless clearly stated otherwise.\\

The resonance condition is satisfied when the ratio of the polar and radial frequency is 3/2: $\omega_{\theta 0}/\omega_{r 0} = 1.5$ at a given $\tau=\tau_0$. However, the adiabatic approximation does not break down at the specific instant of time $\tau=\tau_0$ where $\omega_{\theta 0}/\omega_{r 0} = 3/2$, but in the neighbourhood of the resonance. The question is where?

Due to the symmetry of the phase around $\tau _0$, we assume that the start of the resonance, $\taustart$, is half the resonance duration before the resonance condition is satisfied at $\tau_0$, i.e., $\taustart = \tau_0 - \taures/2$.
Since we do not know $\tau_0$ beforehand, we use the following approach to trigger the start of the resonance in our model. If we define a threshold function as $\xi = \omega_{\theta }/\omega_{r } - 1.5$, which is zero at $\tau_0$, it is possible to find the value of this function at $\taustart = \tau_0 - \taures/2$ such that the total duration of the resonance is $\taures$. To do this, we expand the threshold function around $\tau_0$ up to the first order
\bea
\xi(\tau) 
&\approx 0 + \eval{
\qty(
\frac{\dot{\omega} _\theta}{\omega_r } - \frac{\omega_\theta}{\omega_r ^2} \dot{\omega}_r  
)
}_{\tau_0} (\tau - \tau_0) \\
&\approx \frac{2\dot{\omega} _{\theta 0} - 3 \dot{\omega}_{r0} }{2\omega_{r0} }  (\tau - \tau_0) 
= 2 \pi \frac{\tau - \tau_0}{\taures ^2 \omega_{r0}} \, .
\eea
Thus, the threshold function at $\taustart = \tau_0 - \taures/2$ is $\xi^* = \xi(\tau_0 - \taures/2) = -\pi /(\taures \omega_{r0})$. 
Assuming that $\taures$ and $\omega_{r0}$ are approximately constants in the neighborhood of the resonance, we can compute $\xi = \omega_\theta /\omega_r -1.5$ and $\xi^* = -\pi/(\taures \omega_{r0})$ and will trigger the start of the resonance when $\xi > \xi ^* $. 
This is a universal choice which is independent of the specific EMRI system since the threshold is dimension-free. We verified that by using this criterion to turn on the resonance correction, the ratio 3/2 was reached exactly at $\tau _0$, as expected.

So, if $ |\omega_{\theta 0}/\omega_{r 0} - 1.5|< |\xi^*|$ and if the orbit is non-circular and non-equatorial, we modify the flux evolution of the Numerical Kludge by implementing an Effective Resonance in coordinate time:
\bea
\label{flux_effective_res}
\dv{J}{t} &= f_{\text{NK}} \, \qty{ 1+ 
\mathcal{C} \,  w(t) } \, , 
\eea
where we define the resonance coefficients $\mathcal{C} = (\mathcal{E}, \mathcal{L}_z, \mathcal{Q})$ as the fractional change in flux on resonance, and the ``impulse" function $w$ as
\bea
\label{eq:impulse_fct}
w(t) = 
\begin{cases}
  \frac{1 +\cos \qty[4 \pi \qty ( \frac{t - \tau _0}{\taures})^2]}{\int _0 ^1 1 + \cos \qty[4 \pi x^2] \, \dd x}
  &\quad t \in[\taustart, \taustart +\taures]\\
  0 &\quad \text{elsewhere}
\end{cases}
\, ,
\eea
to represent the evolution of the second term in the flux equation (\ref{eq:flux-resonant}) due to the resonance.

\begin{figure}
    \begin{center}
    \textbf{Evolution of the constants of motion through a 3:2 resonance}
      \includegraphics[width=\linewidth]{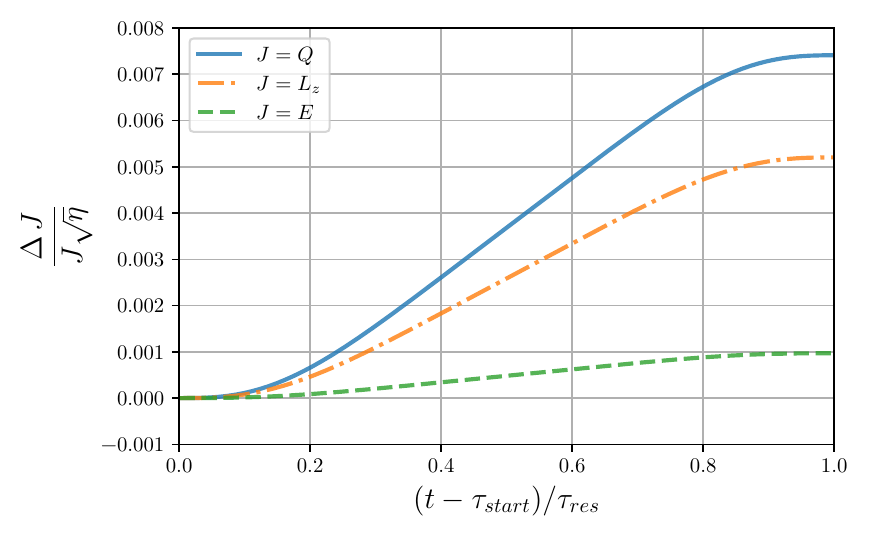}
     \end{center}
  \caption{Transient orbital resonances modify the evolution of the constants of motion and lead to a deviation $\Delta J$ from the standard adiabatic evolution. Here we show the relative changes in the constants of motion from the start to the end of a 3:2 resonance in an EMRI system  with resonance coefficients $\mathcal{C} = (-0.01,-0.01,-0.01)$ and parameters $\eta = 3 \times 10^{-6}, M = 10 ^6 \text{M}_\odot, p/M = 8.55,  e = 0.7 , \iota = 1.20, a = 0.95 $.}
    \label{fig:constants_time_ev}
  \end{figure}

In Figure (\ref{fig:constants_time_ev}) it is shown how the effective model changes the evolution of $J$ and implements a smooth deviation in the constants of motion during the resonance. The size of the changes are different for each of the constants of motion and they are of order $\sim 0.5 \%$ by the end of the resonance.

\subsection{Resonance Coefficients}
Determining the resonance coefficients $\mathcal{C}$ as a function of the EMRI parameters is possible, but is very challenging.
The value of $\mathcal{C}$ for a specific trajectory can be estimated by using Eq.~(\ref{eq:flux-resonant}) following the procedure described in \cite{flanaganResonantlyEnhancedDiminished2014}.
However, this procedure is currently too computationally expensive for our purposes.
As far as we are aware, the only available resonance coefficient estimates are given in \cite{flanaganResonantlyEnhancedDiminished2014} and their largest values reach $\sim 0.01$.

We can check that the Effective Resonance Model reproduces an evolution of the orbital constants and phases similar to that one given in \cite{flanaganTransientResonancesInspirals2012}.
There, Flanagan and Hinderer found a dephasing of $15$ cycles by the end of the inspiral and a change in constants of $ \Delta J/(J \sqrt{\eta}) \approx 0.02 \,- \,0.10$.
The Effective Resonance Model reproduces also the same order of magnitude deviations if we set $\mathcal{C} = (-0.01,-0.01,-0.01)$.

The sign of $\mathcal{C}$ can be positive or negative according to the phase $\Psi_0$ at which the compact object reaches the resonance.
Determining the correct phase $\Psi _0$ at resonance can be problematic since it would require the waveform model to be perfectly in phase from the beginning of the inspiral up to the start of the resonance.
Therefore the resonance coefficients $\mathcal{C}$ are considered as parameters of our Effective Resonance Model.
Once we have predictions from the GSF, the resonance coefficients could be used as a new way to test GR.
For the time being we take as fiducial values $\mathcal{C} = (-0.01,-0.01,-0.01)$.

\subsection{Impulse function}
The ``impulse" function $w(t; \taures, \taustart)$ expresses the rising and the disappearing of the additional term during the resonance, and its functional form has been chosen such that:
\begin{itemize}
  \item[$(i)$] it is symmetric with respect to $\tau_0 = \taustart + \taures/2$ ;
  \item[$(ii)$] it is smooth, positive and normalized to $\taures$ ;
  \item[$(iii)$] it rises from zero with zero slope $w'(t = \taustart) = 0$, and symmetrically it decays to zero with zero slope $w'(t = \taustart + \taures) = 0$ ;
  \item[$(iv)$] it resembles the functional form of the correction in Eq. (\ref{eq:flux-resonant}): $\exp(i x^2) \sim \cos x^2$ ;
\end{itemize}
The first condition $(i)$ is imposed to preserve the symmetry with respect to $\tau_0$ as expressed in equation (\ref{eq:flux-resonant}). The impulse function is normalized to $\taures$ because the total change in the constants of motion is proportional to the resonance duration.
Condition $(iii)$ and $(ii)$ avoid abrupt behaviors in the evolution of the constants.
Conditions $(i,ii,iii)$ are chosen according to the study of the semi-analytic solution of resonances given in~\cite{mihaylovTransitionEMRIsResonance2017}.
Condition $(iv)$ is suggested by the functional form of the equations regulating the resonances.

Since condition $(iv)$ is guided by our intuition, we test that the final waveform is not affected by this choice. 
Therefore, we compute the mismatch $\mathcal{M}$  between one-year EMRI waveforms with different impulse functions in the Effective Resonance Model (initial conditions given in \citep{flanaganTransientResonancesInspirals2012}). In Table (\ref{tab:overlaps_impulse_fct}) we report the mismatch between the waveforms generated with our standard (S) impulse function $w$ of eq.(\ref{eq:impulse_fct}) and four different impulse functions: Blackman–Nuttall window (B-N), Blackman–Harris window (B-H), Tukey window (T) and Nuttall window(N) \citep{1455106,6768513,SASPWEB2011}.
We also report the mismatches for different resonance coefficients $\mathcal{C}$.
\begin{table}[h]
    \centering
    \begin{tabular}{cccc}
        \hline
        $\mathcal{C}$ & $-0.01$  & $-0.005$ & $-0.0025$   \\
        \hline
        $\mathcal{M}(h_\text{S}, h_{\text{B-N}} ) $ & $1.67 \times 10^{-4}$  &  $4.18 \times 10^{-5}$ & $1.046 \times 10^{-5}$\\
        $\mathcal{M}(h_\text{S}, h_{\text{B-H}} )$ & $1.74 \times 10^{-4}$ & $4.38 \times 10^{-5}$ & $1.094 \times 10^{-5}$\\
        $\mathcal{M}(h_\text{S}, h_{\text{T}} )$ &  $4.16 \times 10^{-5}$ & $1.04\times 10^{-5}$ &$2.60\times 10^{-6}$  \\ 
        $\mathcal{M}(h_\text{S}, h_{\text{N}} )$ & $1.79 \times 10^{-4}$ & $4.49 \times 10^{-5}$ & $1.12\times 10^{-5}$\\
        \hline
    \end{tabular}
    \caption{Mismatch $\mathcal{M}$ between one-year waveforms generated with different impulse functions and different resonance coefficients $\mathcal{C}$. Notice that if we halve the value of the resonance coefficient, the mismatch gets four times smaller. Since the two polarizations show the same mismatch, we report the results only for $h_+$.}
    \label{tab:overlaps_impulse_fct}
    \end{table}
As shown from Table~\ref{tab:overlaps_impulse_fct}, the mismatches are particularly low for all the considered cases, $\sim 10^{-4}-10^{-6}$.
Interestingly, the value of the mismatch is quadrupled if the resonance coefficients is doubled. An explanation for this behavior will be provided in Sec.~\ref{subsec:mismatch}. 
Since the largest expected resonance coefficients $\mathcal{C} = (-0.01,-0.01,-0.01)$ lead to a mismatch of order $10^{-4}$, there is little difference between using different impulse functions.

We conclude that the functional form of the impulse function does not matter too much, as long as the phase immediately after resonance is correctly modelled.
In fact, we do not aim to exactly describe the evolution of the inspiral during the resonance, but to correctly account for their effects afterward, which is where large dephasings can accumulate. As long as we are back in phase after the resonance, and the resonance duration is relatively short, we can use the Effective Resonance Model to match astrophysical resonant EMRIs without significant loss of signal-to-noise ratio. We will use the impulse function of Eq.~(\ref{eq:impulse_fct}) for the rest of this work

Our aim is to study how resonances influence the gravitational wave signals from EMRIs and how parameter estimation could be affected.
Even though we are using a phenomenological model and we do not use the exact value of the resonance coefficients, this model is a reasonable starting point to understand at which order of magnitude resonances are going to affect the gravitational waves and EMRI parameter estimation. 
In addition, it provides a new tool for observing and measuring resonances.
The Numerical Kludge combined with this phenomenological model has the advantage that it is fast enough to scan a large EMRI parameter space and to capture the crucial physics of resonances and EMRI evolution.

\section{Results}
\label{sec:results}

\subsection{Dephasing caused by 3:2 resonances}

\begin{figure}
  \begin{center}
    \textbf{Dephasing over the parameter space}
    \includegraphics[width=\linewidth]{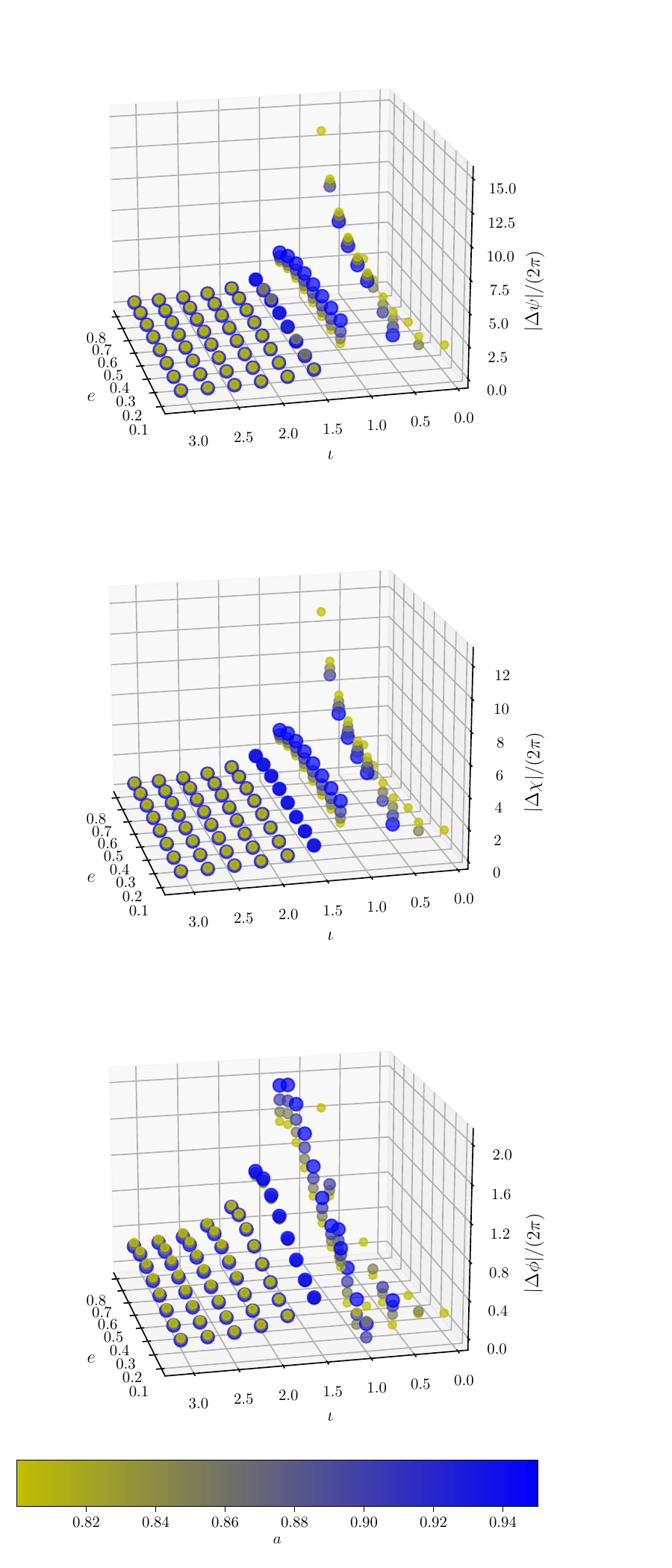}
    \end{center}
\caption{
The dephasing between a resonant and non-resonant orbit in the phases $\psi$ (top panel), $\chi$ (middle panel) and $\phi$ (bottom panel).
The dephasings are computed for EMRI systems with parameters $\eta=10^{-5}, M = 10^6 \, \text{M}_\odot$ and resonance coefficients $\mathcal{C} = (-0.01,-0.01,-0.01)$.
Since the phases oscillate quickly we take the the average over the last 1000 $s$. Some small variations in the dephasing pattern may arise from this procedure.}
  \label{fig:phases_config}
\end{figure}

The gravitational wave of an EMRI system consists of a superposition of multiple modes. Each mode is characterized by an amplitude and an oscillating sinusoid. The phase evolution of this oscillating term depends on a particular combination of the three Keplerian phases $(\psi, \chi, \phi)$ associated with the EMRI. The overall gravitational wave phase of an EMRI system thus consists of a complex combination of the evolution of the phases $(\psi, \chi, \phi)$.
Resonances cause these phases to depart from the standard adiabatic evolution. In this section we will quantify the size of these dephasings in terms of how many cycles the resonant evolution deviates from from the non-resonant evolution, $(\Delta\psi, \Delta\chi, \Delta\phi)/(2 \pi)$, by the end of the one year observation. 

This analysis will serve as a verification of the Effective Resonance Model, and allow comparison with prior work, but we will also try to quantify the dephasings in a different way. Previous investigations have considered a limited set of EMRI configurations and analysed the dephasings induced by resonances using a Post-Newtonian model for the resonant force~\citep{berryImportanceTransientResonances2016,flanaganTransientResonancesInspirals2012}. 
The Post-Newtonian force model predicts a particular relationship between the parameters of the EMRI at resonance and the size of the flux changes on resonance, which may or may not be a good approximation to the true impact of the GSF.
In the Effective Resonance Model, by contrast, the size of the flux changes is set by hand, and so we can compare the dephasing with these fixed and thus decouple the effect of the size of the flux change from the timing of the resonance within the inspiral. These results can thus be more easily reused once GSF calculations have accurately computed the on-resonance flux changes.
In the following, we will use our fiducial resonance coefficients $\mathcal{C} = (-0.01,-0.01,-0.01)$ to estimate the dephasings. This can be regarded as a worst case scenario, but these dephasings can be readily rescaled to other choices.

By studying which regions of the EMRI parameter space lead to the largest dephasings, we can understand which EMRI configurations are mostly likely to be affected by 3:2 resonances. Thus, we evolve several EMRI systems with initial conditions $e \in [0.1, 0.8]$, $\iota \in [0.1, 3.1]$, $a \in [0.8, 0.95]$, with steps, $\Delta e = 0.1$, $\Delta \iota = 0.3$ and $\Delta a = 0.05$. The initial semi-latus rectum $p/M$ is determined using the Black Hole Perturbation Toolkit \cite{BlackHolePerturbation} to start the inspiral near the resonance, more precisely at $|\xi| = 3/2-1.498$ (see Appendix \ref{app:3to2_loc} for the parameter space location of the 3:2 resonance). 
We limit our analysis to configurations that have an initial periastron $r_p > 5 \,M$ for consistency with the limitations of the NK model.
These parameter ranges are chosen such that they span the range of eccentricities $e$, inclination angle parameters $\iota$ and spins $a$ expected for astrophysical EMRIs~\cite{babakScienceSpacebasedInterferometer2017, mcclintockSpinNearExtremeKerr2006,risalitiRapidlySpinningSupermassive2013,gouEXTREMESPINBLACK2011}. 
We fix the mass ratio and the mass of the central black hole to $\eta = 10^{-5}, M= 10^6\, \text{M}_\odot$, respectively, and we set the initial phases to zero. For most of these configurations, the dephasing was computed over an observation time of one year. For systems with $\iota<1.3$ the location of the resonance in parameter space is such that the systems plunge in less than one year. In those cases the dephasing was calculated at plunge.

The dephasings for these systems are shown in Figure~(\ref{fig:phases_config}). 
They mildly depend on the spin parameter and span a wide range of values from $\sim 0.3$ increasing up to $(\Delta\psi, \Delta\chi, \Delta\phi)/(2 \pi) =(14.8,12.2 ,2.0)$.
The largest shifts due to resonances are located around $\iota \sim 1.2$ as shown in Figure~(\ref{fig:phases_config}).

The dephasing for fixed inspiral length scales like $\Delta\dot{\omega} \, T^2$, where $\Delta\dot{\omega}$ is the change in the frequency derivative that accumulates over the resonance, and $T$ is the time between the resonance and the end of the inspiral.
The change in frequency derivative depends on the difference in the energy flux at resonance, and on the duration of the resonance. 
So, the largest phase shifts will be for systems that have large changes in the flux on resonance and long resonance durations. 
The longest resonance durations are for systems with high eccentricities and high $\iota$, but the biggest phase shifts are for systems with smaller $\iota$.
These latter systems have resonances which occur closer to the central black hole, where the absolute magnitude of the dissipation rate is higher. 
For fixed $T$, this leads to a higher $\Delta J (\taures + \taustart)$ compared to other configurations and, as a consequence, higher phase shifts $(\Delta\psi, \Delta\chi, \Delta\phi)/(2 \pi)$, as we can see from Figure~(\ref{fig:phases_config}). The turnover in the dephasing for configurations with $\iota \lesssim 1.0$ is due to the fact that these systems plunge within one year.  The total time over which the dephasing can accumulate decreases as $\iota$ decreases from $\sim 1.0$ to zero, and this decrease compensates for the increased strength of the resonance effect to give a total dephasing that is decreasing.

Overall we conclude that, for fixed inspiral length, we expect the strongest impact on the waveform from the 3:2 resonances located in the strong field regime, i.e., closer to the central black hole.

\subsection{Mismatch as a function of resonance strength}
\label{subsec:mismatch}
The dephasing due to a resonance accumulates over the inspiral, which means the overlap between a resonant and non-resonant waveform drops over time (or equivalently the mismatch increases). This has been shown to affect the detection of EMRI systems \cite{berryImportanceTransientResonances2016}. The size of the accumulated mismatch will depend also on the strength of the resonance. Here, we study the evolution of the mismatch, $\mathcal{M}$, as a function of the resonance coefficients, $\mathcal{C}$, for different EMRI initial conditions.
By doing this we can assess which values of $\mathcal{C}$ lead to a significant dephasing.

\begin{figure}
    \begin{center}
    \textbf{Mismatch as a function of the resonance strength}\par\medskip
      \includegraphics[width=\linewidth]{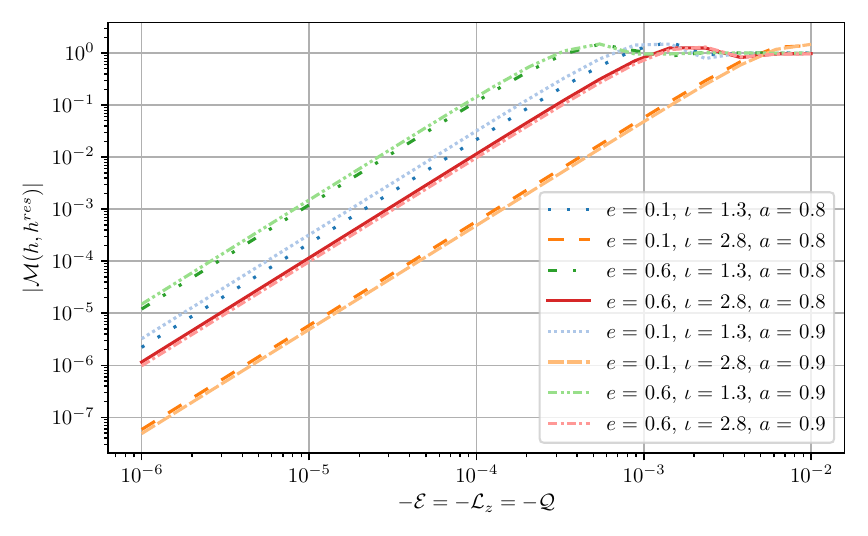}
      \end{center}
  \caption{Mismatch $\mathcal{M}$ of resonant and non-resonant waveforms as a function of the resonance coefficients $\mathcal{C} = (\mathcal{E},\mathcal{L}_z,\mathcal{Q})$. We considered a one-year waveform generated for an EMRI system with $\eta=10^{-5}, M = 10^6 \, \text{M}_\odot$ and different initial conditions. The dashed and solid lines correspond to systems with spin $a=0.9$ and $a=0.8$, respectively. The lines with the same color have the same $\iota$. 
  }
    \label{fig:coef_olap}
  \end{figure}

We consider a one-year inspiral of an EMRI system with $\eta =10^{-5}$ and $M =10^6 \, \text{M}_\odot$. For simplicity we set the three resonance coefficients to be equal $\mathcal{E} = \mathcal{L}_z = \mathcal{Q}$, but have verified that the results are not significantly influenced by this choice. As we increase the size of the resonance coefficients, we calculate the mismatch $\mathcal{M}(h,h^{res})$ between non-resonant and resonant waveforms for the two LISA channels. Since the results of the two channels are basically identical we show the results in Figure~(\ref{fig:coef_olap}) for channel $I$ only.

These results reflect some of the previous findings. The region of the parameter space where the dephasings are larger have a larger mismatch.
In fact, prograde orbits (dash-dot-dotted green and dotted blue lines) show a larger mismatch than retrograde (dashed orange and, solid and dashdotted red lines).
The spin parameter mildly affects the overall behavior.
For higher eccentricity systems, the mismatch begins to increase significantly for smaller absolute values of the resonance coefficients.
We find a direct proportionality between the mismatch $\log|\mathcal{M}|$ and the resonance coefficients $\log\mathcal{C}$ for $\mathcal{C}<3\cdot10^{-4}$. This is due to the fact that the mismatch scales like the dephasing squared, and, as previously discussed, the dephasing at the end of the inspiral scales approximately like $\sim \Delta \dot{\omega}\sim \mathcal{C}$. Therefore the scaling $\log|\mathcal{M}|\propto\log\mathcal{C}$ is not unexpected.

A common criterion for assessing if two waveforms are indistinguishable is that the norm of the waveform difference, $(\delta h | \delta h) < 1$. This criterion was first introduced in~\cite{Flanagan:1997kp} and ~\cite{PhysRevD.71.104016}, but was popularised by~\cite{Lindblom:2008cm}. The mismatch is approximately $(\delta h|\delta h)/(2 \rho^2)$ and so the indistinguishability criterion is satisfied for ${\cal M} \lesssim 1/(2 \rho^2)$, which is approximately $10^{-3}$ for the $\rho \sim 20$ systems considered here. From Figure~(\ref{fig:coef_olap}) we conclude that for resonance coefficients smaller than $\sim 10^{-5}$ the waveforms are indistinguishable, regardless of the choice of system parameters, and so the effects of the resonance can be neglected. Furthermore, we infer that the mismatch strongly depends on the parameter space location of the EMRI system and that the largest mismatches, for a fixed size of the resonance coefficients, occur for high eccentricity, prograde orbits in the small resonance strength regime $|\mathcal{C}|<10^{-3}$.

We have considered a worst case scenario for the mismatch because we did not maximize over the initial time, phases or other model parameters in order to find the best-fit template, as is normally done in GW data analysis.
However, we do not expect the above results to drastically change with this maximization, as already demonstrated in~\citep{berryImportanceTransientResonances2016}.

\subsection{Systematic errors}

\begin{figure}
    \begin{center}
    \textbf{Resonance induced biases of intrinsic parameters}\par\medskip
        \includegraphics[width=\linewidth]{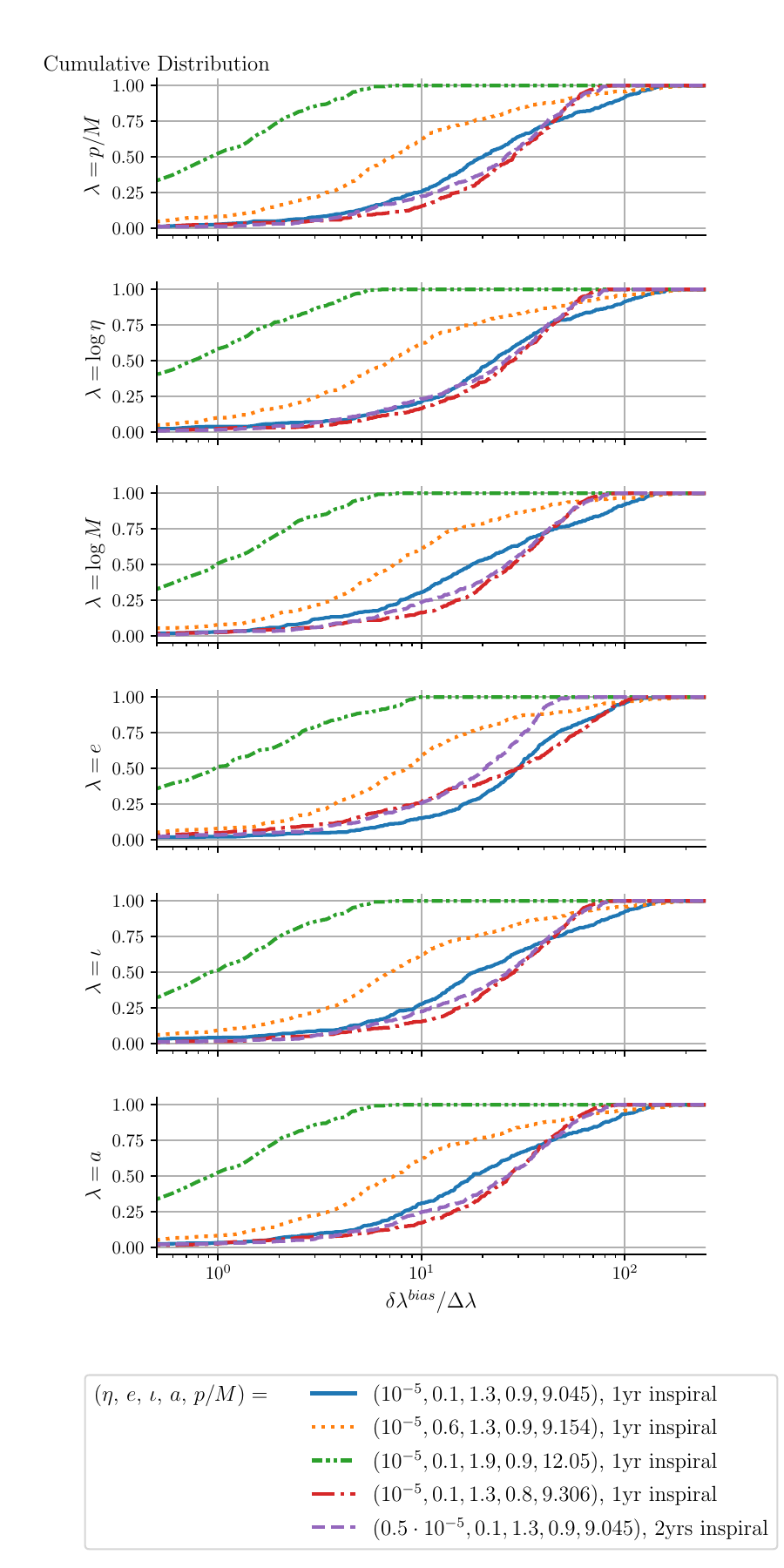}
      \end{center}
  \caption{Cumulative distribution of the biases (or systematic errors) induced by resonances for different intrinsic parameters and configurations. The ratio $\delta \lambda ^{bias}/\Delta \lambda$ between the size of the systematic and statistical errors is calculated for one-year inspiral by randomly drawing resonance coefficients. For $\delta \lambda ^{bias}/\Delta \lambda>1$ the bias induced by inaccurate modeling is larger than the bias induced by the typical noise fluctuations. This figure shows how the bias is distributed over many resonance strength realizations (resonance coefficient realizations). For the fiducial EMRI configuration (solid blue line), more than $95\%$ of the biases are larger than two sigma, i.e. the cumulative distribution reaches $\approx 0.05$ at $\delta \lambda ^{bias}/\Delta \lambda=2$.
  }
    \label{fig:bias_singleEMRI}
  \end{figure}
The use of an approximate waveform model causes biases in parameter estimation. Therefore, the scientific potential of EMRI systems can be undermined if the systematic error $\delta \lambda ^{bias}$ induced by inaccurate waveform modeling is larger than the size of the typical statistical error (Eq.~\ref{eq:measurement_err}) \cite{vallisneriStealthBiasGravitationalWave2013}. Statistical errors arise due to noise fluctuations and are therefore unavoidable, but are accounted for in the width of posterior distributions on parameters. Requiring the systematic error to be smaller than the statistical error ensures that the true parameters remain consistent with the posterior for single events, although errors can accumulate for populations~\cite{2015PhRvD..91l4062G}. The statistical error can be estimated as the square root of the diagonal elements of the inverse of the Fisher matrix $\Delta \lambda$ and so we are interested in the ratio
\bea
\qty(\frac{\delta \lambda ^{bias}}{\Delta \lambda})^i &=  \frac{
\qty|
    \qty[\qty(\Gamma _I + \Gamma _{II} )^{-1}]^{i} \, _k \, \qty(b_I + b_{II} )^k
    |
}{
    \sqrt{
    \qty[\qty(\Gamma _I + \Gamma _{II} )^{-1}]^{ii}
    }
}\\
b^k _{ch} &=  \eval{
\braket{\partial_k h_{ch} }{h_t ^{ch} - h^{ch} _m}
}_{\vb*{\lambda}=\vb*{\lambda}_0} \qquad ch = I, \, II
\, ,
\eea
where there is no sum over the index $i$, $h_m$ and $h_t$ refer to the approximate and true waveforms respectively, and each $i$ refers to one of the considered parameters $\vb*{\lambda} = (p/M,\log\eta,\log M, e,\iota)$. The subscript $I, II$ on the Fisher matrices mean that they have been calculated for the respective detector channels. Here we investigate the size of the errors that arise from ignoring the presence of resonances in the waveform model. The approximate waveform, $h_m$, will be the NK waveform without resonance and the true waveform, $h_t$, will be the waveform produced with the Effective Resonance Model. 

We consider a fiducial EMRI configuration $\eta =10^{-5}, M = 10^6 \, \text{M}_\odot, e=0.1, \iota =1.3, p/M = 9.0447, a = 0.9$. The resonance coefficients, $\mathcal{C}$, are in principle a function of the waveform parameters and can be computed using the GSF. However, such calculations are beyond the scope of this paper and so we draw random values for the resonance coefficients from a uniform distribution ($\mathcal{E}\sim\mathcal{L}_z\sim\mathcal{Q}\sim U[-0.01,0.01]$, where each resonance coefficient is independently sampled). Then, we look at the distribution of biases over repeated random draws of this kind.
The domain of the uniform distribution is chosen to cover all the expected resonance coefficient values. To explore the impact of the other waveform parameters we repeat this procedure for other EMRI systems which differ from the fiducial system by the change of one parameter at a time. For each system, the semi-latus rectum was adjusted to ensure the 3:2 resonance was encountered close to the start of the inspiral, i.e., at the start of the inspiral we had $|\xi| = 3/2 -1.498$. In all cases the EMRI did not plunge within the one year observation time used. 

We show the cumulative distribution of the biases 
for the various EMRI systems in Figure (\ref{fig:bias_singleEMRI}).
The cumulative distribution of $\delta \lambda ^{bias}/\Delta \lambda$ for the fiducial configuration (blue solid lines) shows that more than $95\%$ of the biases are more than twice the statistical uncertainty, $\Delta \lambda$, (``two sigma away''). Reducing the spin (red dashdotted lines) or the mass ratio $\eta$ (violet dashed lines) does not significantly change the cumulative distribution.
For these configurations we conclude that EMRI waveform models need to account for resonances. For the EMRI systems with higher eccentricity (orange dotted lines) and larger $\iota$ (green dash-dot-dotted lines), the proportion of systems with biases above 2 sigma is lower, but these proportions still exceed $\sim 85\%$ and $\sim 25\%$ respectively. These results are consistent with the mismatch analysis presented earlier, where we found typically lower mismatches for more retrograde systems. 

Overall, the conclusion from Figure~(\ref{fig:bias_singleEMRI}), is that it is important to use waveform models that include resonances when performing EMRI parameter estimation. The biases will be less severe for retrograde and higher eccentricity systems, although even there the biases are not negligible and will continue to accumulate if the observation time is increased beyond one year.

The bias estimates here were calculated using the approximate formalism described in~\citep{vallisneriStealthBiasGravitationalWave2013} and its validity will degrade for large biases, $\delta \lambda ^{bias}/\Delta \lambda \gtrsim 10$. This does not invalidate our findings but it is likely to affect the reliability of our estimate of the tails of the cumulative distribution of biases. The tails could be more accurately resolved by computing posterior distributions for each configuration. This is computationally expensive and so is beyond the scope of the current work, but we do not expect the conclusions to be significantly different in such an analysis.

\subsection{Measurability of Resonances}
\begin{figure}
    \begin{center}
    \textbf{Measurability of intrinsic parameters and resonances}\par\medskip
        \includegraphics[width=\linewidth]{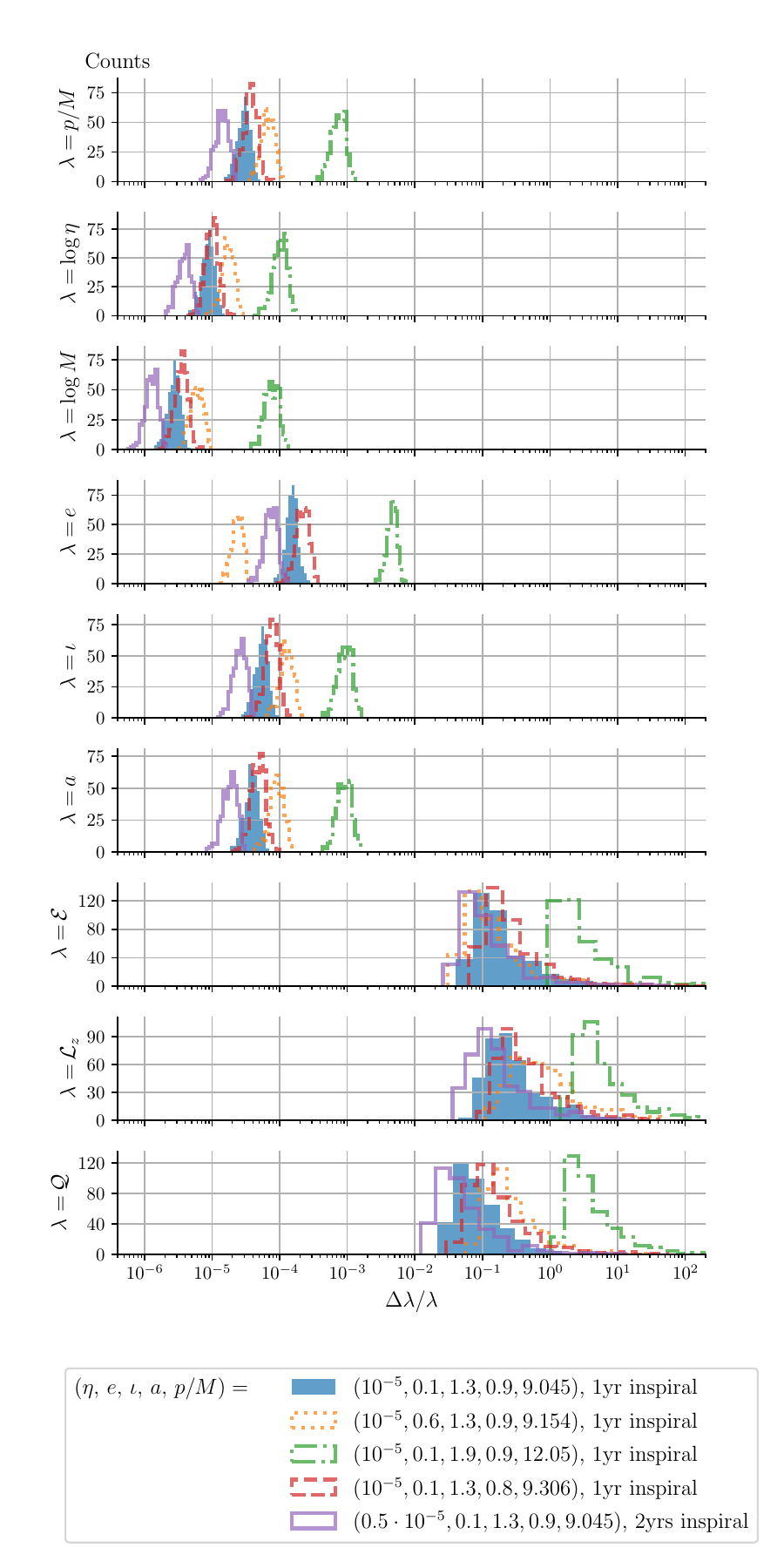}
      \end{center}
  \caption{Distribution of the fractional measurement precision $\Delta \lambda / \lambda$ of intrinsic parameters and resonance coefficients, $\mathcal{E}, \mathcal{L}_z, \mathcal{Q}$. The counts represent the fraction of times a given EMRI system has a particular measurement precision over random draws of the resonance coefficients.
  All the signals have been normalized to SNR $=20$. Parameters can be considered measurable if $\Delta \lambda/\lambda <1$.
  This figure shows how the measurement precision is distributed over many resonance strength realizations (resonance coefficient realizations).
  For the fiducial EMRI configuration (filled blue histogram), the resonance coefficients $\mathcal{E}, \mathcal{L}_z, \mathcal{Q}$ are determined in median with relative precision $(0.15,0.23,0.07)$.
  }
    \label{fig:fish_singleEMRI}
  \end{figure}

We now investigate two things: the measurability of resonances and how the inclusion of the resonance coefficients $\mathcal{C}$ as unknown model parameters change the measurement precision of other intrinsic parameters.
This is assessed using the Fisher information matrix.
If $\Delta \lambda/\lambda<1 $ then the parameter is measurable.
For details about the validation and checks of our Fisher matrix calculations we refer the reader to Appendix~\ref{app:fish_validation}.

As before, we consider a fiducial EMRI configuration and we calculate the relative measurement precision $\Delta \lambda /\lambda$ for uniformly randomly drawn resonance coefficients.
We repeat this for the same set of reference systems considered in the previous section, which vary one parameter at a time away from the fiducial model. 
All the waveforms have been normalized to SNR $ = 20$ and we compute measurement precisions for the parameters $(p/M, \eta,  M, e,\iota, a, \mathcal{E}, \mathcal{L}_z, \mathcal{Q})$.
We show the results for the distribution of parameter measurement precision in Figure~(\ref{fig:fish_singleEMRI}).

Measurement precisions, $\Delta \lambda / \lambda$, for the intrinsic parameters $(p/M, \eta,  M, e,\iota, a)$ are at the sub-percent level in all cases, and vary by a factor of a few across the distribution of resonance coefficient values. This shows that the inclusion of the resonance coefficients as additional parameters does not significantly degrade the measurement precision for the intrinsic parameters. In addition, we find that these precision estimates are in agreement with the typical values found in \citep{babakScienceSpacebasedInterferometer2017}.

It is clear from Figure~(\ref{fig:fish_singleEMRI}) that the resonance coefficients cannot be constrained for the high $\iota$ configuration (green dashdotted). In fact as previously noticed, the parameter space location of the 3:2 resonance for high $\iota$ is particularly far ($p/M\sim 12$) from the strong field regime and this leads to smaller dephasings with respect to the other configurations. Since the time to plunge for this configuration (green dashdotted) is three years, we expect that if we had evolved such an EMRI system for a longer time, the measurement precision of the resonance coefficients for this system would improve.

On the other hand, the most precise measurement estimates are obtained for the smaller mass ratio system (violet solid).
This is probably due to the choice of increasing the observation time for this system. While that was done to ensure the amount of inspiral was comparable, it means that twice as many waveform cycles are also observed so we expect to measure parameters somewhat better. 
We find that for this EMRI configuration the resonance coefficients $\mathcal{E}, \mathcal{L}_z, \mathcal{Q}$ are determined with relative precision better than $0.1$, respectively, $54\%, 35\%, 81\%$ of the time.
Here, we report the median relative precision of the three resonance coefficients, $(\text{Md}[\mathcal{E}],\text{Md}[ \mathcal{L}_z], \text{Md}[\mathcal{Q}])$. The medians are computed for the distributions of each of the three resonance coefficients and are $(0.15,0.23,0.07)$ for the fiducial configuration (blue filled), $(0.11,0.68,0.21)$ for the more eccentric one (orange dotted), and $(0.20,0.35,0.13)$  for the low spin one (red dashed).

\section{Discussion and future outlook}
Extreme Mass Ratio Inspirals are modeled using perturbation theory in the small mass ratio $\eta$.
The effects of transient orbital resonances on the EMRI phase evolution scale as $\sim \eta ^{-1/2}$, therefore contributing more than post-adiabatic corrections which scale as $\sim \eta^{0}$.
It has been shown how resonances affect detection \citep{berryImportanceTransientResonances2016}. However, the impact of these phenomena on parameter estimation had not previously been investigated. 

In this work we have explored this question by implementing a phenomenological model for transient orbital resonances: the Effective Resonance Model.
This model allows us to efficiently study a wide variety of EMRI systems avoiding expensive gravitational self-force calculations, but at the cost of introducing three additional model parameters: the resonance coefficients $\mathcal{C}=(\mathcal{E},\mathcal{L}_z, \mathcal{Q})$.
These encode the relative flux changes for each of the constants of motion that occur during resonance, and, therefore represent the resonance strength. Since the largest estimated resonance coefficients in the literature are of order $\sim 0.01$, we adopt the fiducial values $\mathcal{C}=(-0.01,-0.01,-0.01)$ to study the dephasings induced by 3:2 resonances over the parameter space. 
For a one-year EMRI inspiral with parameters $\eta = 10^{-5}, M=10^6\, \text{M}_\odot$, we find that the maximum dephasings amount to $(14.8,12.2 ,2.0)$ cycles for $(\psi, \chi, \phi)$ respectively, and occur for prograde orbits.
This is due to the fact that the 3:2 resonances of these configurations are located closer to the central MBH where the absolute magnitude of the fluxes is higher.

We have also investigated how the mismatch between a resonant and non-resonant waveform depends on the resonance coefficients, finding a proportionality between the logarithm of the mismatch and the logarithm of the resonance coefficients. The study of the mismatch was consistent with the behavior of the dephasings over the parameter space. In addition, it revealed that more eccentric systems might be more affected by resonances in the small resonance strength regime $|\mathcal{C}|<10^{-3}$. We also inferred that resonance coefficients with absolute values less than $\sim 10^{-5}$ lead to waveforms that are indistinguishable from the waveforms without resonances and hence there is no measurable effect in the phase evolution.

We conducted a study of the systematic errors that would arise in estimates of the intrinsic parameters from neglecting resonances. We found that for the considered EMRI configurations the biases can reach values up to a dozen times larger than the statistical errors arising from noise fluctuations. Over all the resonant coefficient realizations, we find that more than $95\%$ of the biases are larger than 2 sigma, for our fiducial EMRI configuration.
This suggests that parameter estimates of resonant EMRIs are likely to be biased if resonances are not taken into account in models used for parameter estimation. We also expect an increase in the parameter bias for longer signals.

Finally, we used the Effective Resonance Model to assess the measurability of parameters with resonant EMRIs. The EMRI systems which are mostly affected by resonances provide measurements of the resonance coefficients with a relative precision ranging from $0.07-0.68$ (median values).

The Effective Resonance Model presented here has a number of potential uses, beyond being used to scope out the impact of resonances as done in the current paper. A model similar to this could be used in analysis of LISA data to match EMRI signals over resonances, mitigating the induced biases without requiring expensive GSF calculations of the resonance coefficients. 
Furthermore, due to its general implementation, the Effective Resonance Model can also be adapted to represent tidal resonances \citep{bongaTidalResonanceExtreme2019}, which have a similar form to transient orbital resonances but are caused by the tidal perturbation of a third object. Finally, such a model can be used to measure resonance coefficients in EMRI observations. Comparing the measured values to the self-force prediction of the resonance coefficients would provide a test of general relativity.

There are a number of ways in which this work could be extended. We have explored only 3:2 resonances here, so it would be interesting also to analyse higher order resonances, or systems that pass through multiple strong resonances. In addition, our parameter estimation and bias results were based on the linear signal approximation, and it would be useful to verify these using full Bayesian posterior calculations. Finally, once full GSF waveforms with resonances are available, the ability of the Effective Resonance Model to detect and characterise such systems should be assessed.

While the development of an accurate self-force waveform model will be necessary to have more precise and quantitative results, the Effective Resonance Model has provided a first step towards an understanding of the impact of transient orbital resonances on parameter estimation.

\section*{Acknowledgments}
We thank Matthias Bartelmann, Ollie Burke, Maarten van de Meent, Michael Katz, Vojtech Witzany and Béatrice Bonga for their useful suggestions and discussions.
This work makes use of the Black Hole Perturbation Toolkit~\citep{BlackHolePerturbation}.
\section*{Appendix}

\subsection{Data Analysis Caveats}
\label{app:data_analysis}

The Numerical Kludge produces approximate numerical waveforms expressed as an array of data points spaced by a constant time sampling interval, $\Delta t$. 
Since we are not dealing with continuous and analytical signals, we are going to briefly discuss different caveats that can occur when calculating Fourier transforms and derivatives.

The choice of a too large sampling interval can lead to aliasing and it imposes a limit on the maximum resolvable frequency $f_{\text{max}} = 1/(2 \Delta t)$ as prescribed by the Nyquist-Shannon sampling theorem. On the other hand, a too small sampling interval requires a higher computational cost.
Throughout this work we fixed $\Delta t = 10 $s, which is sufficient to resolve the frequencies accessible to the LISA sensitivity band without unduly increasing the computational time.

The waveform derivatives $\partial_j h$ are calculated numerically using the five-point stencil formula to ensure a high stability of the derivatives and a numerical error that scales at fourth order in the derivative spacing. In order to avoid spectral leakage and to sample over the minimum possible frequency range we apply a Tukey window function with shape parameter $0.05$ to every waveform and then zero-pad them to exploit the efficiency of the Fast Fourier Transform  \citep{2020SciPy-NMeth}.

\subsection{Fisher matrix validation}
\label{app:fish_validation}

We make use of the following result, valid for high SNR in the linear signal approximation regime, to check the correctness of the Fisher matrix results and numerical derivative approximations:
\bea
\label{eq:overlap_perturbed}
\mathcal{O} \qty( h(\vb*{\lambda + \gamma}), h(\vb*{\lambda})  ) =
1 - \frac{1}{2 \, \rho ^2} \gamma ^i \Gamma _{ij} \gamma ^j + O(\rho^{-4}) \, .
\eea
The choice of the perturbation $\vb*{\gamma}$ must be small enough to respect the linear signal approximations but also big enough to test the $1\sigma$ region. We follow a similar approach to those described in \cite{vallisneriUseAbuseFisher2008,Burke_2020}, taking $\vb*{\gamma}$ to be a linear combination of the eigenvectors, $v^i _{A}$, and corresponding eigenvalues, $w_A$, of the Fisher matrix, $\Gamma$, where the index $A$ runs over the $N$ different eigenvectors.
We draw coefficients $c_A$ from an $N$ dimensional unit sphere and we define $\vb*{\gamma}$ as follows
\bea
\gamma ^i = \sum _A c_A \,  \frac{v^i _A}{\sqrt{w_A}} \, .
\eea
For every computed Fisher matrix, we calculate separately the left hand side and right hand side, neglecting terms of order $O(\rho ^{-4})$ and higher in Eq.~(\ref{eq:overlap_perturbed}). By taking the ratio between the two sides, we find a maximum deviation of order~$10^{-5}$.
This is a way of checking both the validity of the Fisher matrix and the linear signal approximation.

The Fisher matrices encountered in EMRI data analysis are known for having a high condition number, $\kappa =\text{max}(w_A)/\text{min}(w_A)$, which is the ratio between the maximum and minimum eigenvalue.
A small perturbation in a Fisher matrix with high condition number $\kappa $ can be amplified by a factor of $\kappa$ in the inversion.
Therefore, if we have inaccuracies in our Fisher matrix, $\Gamma$, for instance due to our numerical derivative approximation, they can lead to a wrong value of $\Delta \lambda ^i= \sqrt{(\Gamma ^{-1})^{ii}}$, invalidating our measurability conclusions.
We perturb each element of every Fisher matrix with a deviation matrix $F^{ij}$, where each element is drawn from a uniform distribution $U[-10^{-3},10^{-3}]$, and we calculate
$$
\max_{\text{all configurations}} \qty{ \max_{ij} \qty[\frac{\qty(\qty(\Gamma + F)^{-1} - \Gamma^{-1})^{ij}  }{ (\Gamma^{-1})^{ij} }] } = 0.08 \, .
$$
This confirmed the stability of our Fisher matrix results.

\subsection{Parameter space location of 3:2 resonances}
\label{app:3to2_loc}
\begin{figure}[h]
  \begin{center}
    \textbf{Parameter space location of 3:2 resonances}
    \includegraphics[width=\linewidth]{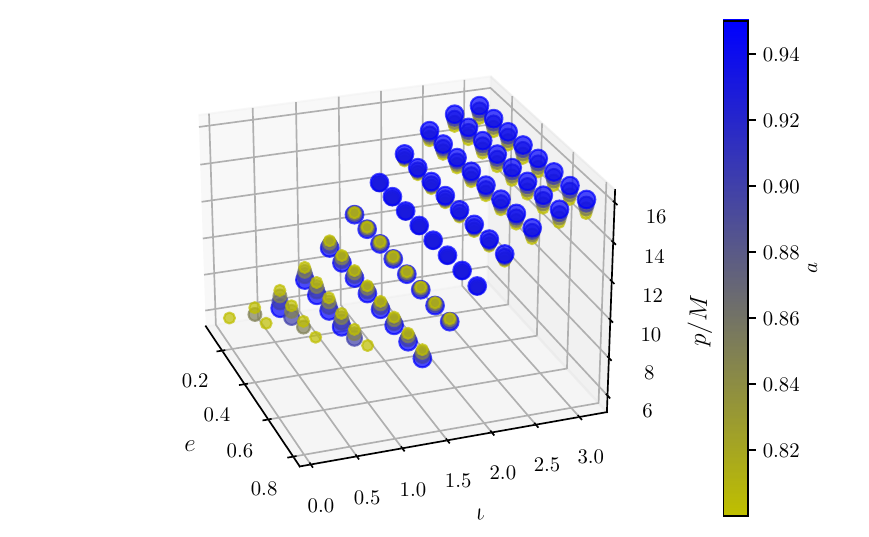}
    \end{center}
\caption{Parameter space location of 3:2 transient orbital resonances.}
  \label{fig:param_space_3to2_loc}
\end{figure}

In Figure~(\ref{fig:param_space_3to2_loc}) we show the parameter space location of 3:2 resonances, calculated using the Black Hole Perturbation Toolkit \citep{BlackHolePerturbation}. The values of $p/M,e,\iota,a$ just before resonance can be calculated by searching the roots of $\omega_\theta /\omega_r = 1.498$. These values are used as initial conditions to evolve EMRI system through resonances.

\bibliography{resonance_paper.bib}

\begin{thebibliography}{10}

\bibitem{theligoscientificcollaborationObservationGravitationalWaves2016}
T.~L.~S. Collaboration and {the Virgo Collaboration}.
\newblock Observation of {{Gravitational Waves}} from a {{Binary Black Hole
  Merger}}.
\newblock {\em Phys. Rev. Lett.}, 116(6):061102, February 2016.

\bibitem{taylorNewTestGeneral1982}
J.~H. Taylor and J.~M. Weisberg.
\newblock A new test of general relativity - {{Gravitational}} radiation and
  the binary pulsar {{PSR}} 1913+16.
\newblock {\em The Astrophysical Journal}, 253:908--920, February 1982.

\bibitem{hulseDiscoveryPulsarBinary1975}
R.~A. Hulse and J.~H. Taylor.
\newblock Discovery of a pulsar in a binary system.
\newblock {\em The Astrophysical Journal Letters}, 195:L51--L53, January 1975.

\bibitem{aasiAdvancedLIGO2015}
a.~J. Aasi et al.
\newblock Advanced {{LIGO}}.
\newblock {\em Class. Quantum Grav.}, 32(7):074001, March 2015.

\bibitem{acerneseAdvancedVirgoSecondgeneration2014a}
F.~Acernese et al.
\newblock Advanced {{Virgo}}: A second-generation interferometric gravitational
  wave detector.
\newblock {\em Class. Quantum Grav.}, 32(2):024001, December 2014.

\bibitem{nitz2OGCOpenGravitationalwave2020}
A.~H. Nitz et al.
\newblock 2-{{OGC}}: {{Open Gravitational}}-wave {{Catalog}} of {{Binary
  Mergers}} from {{Analysis}} of {{Public Advanced LIGO}} and {{Virgo Data}}.
\newblock {\em ApJ}, 891(2):123, March 2020.

\bibitem{theligoscientificcollaborationGW190425ObservationCompact2020}
T.~L.~S. Collaboration et al.
\newblock {{GW190425}}: {{Observation}} of a {{Compact Binary Coalescence}}
  with {{Total Mass}} $ \sim 3.4 \text{M}_\odot$.
\newblock {\em ApJ}, 892(1):L3, March 2020.

\bibitem{nitz1OGCFirstOpen2019}
A.~H. Nitz et al.
\newblock 1-{{OGC}}: {{The First Open Gravitational}}-wave {{Catalog}} of
  {{Binary Mergers}} from {{Analysis}} of {{Public Advanced LIGO Data}}.
\newblock {\em ApJ}, 872(2):195, February 2019.

\bibitem{ligoscientificcollaborationandvirgocollaborationGW170817ObservationGravitational2017}
{LIGO Scientific Collaboration and Virgo Collaboration} et al.
\newblock {{GW170817}}: {{Observation}} of {{Gravitational Waves}} from a
  {{Binary Neutron Star Inspiral}}.
\newblock {\em Phys. Rev. Lett.}, 119(16):161101, October 2017.

\bibitem{ligoscientificcollaborationandvirgocollaborationGW170814ThreeDetectorObservation2017}
{LIGO Scientific Collaboration and Virgo Collaboration} et al.
\newblock {{GW170814}}: {{A Three}}-{{Detector Observation}} of {{Gravitational
  Waves}} from a {{Binary Black Hole Coalescence}}.
\newblock {\em Phys. Rev. Lett.}, 119(14):141101, October 2017.

\bibitem{ligoscientificcollaborationandvirgocollaborationGWTC1GravitationalWaveTransient2019}
{LIGO Scientific Collaboration and Virgo Collaboration} et al.
\newblock {{GWTC}}-1: {{A Gravitational}}-{{Wave Transient Catalog}} of
  {{Compact Binary Mergers Observed}} by {{LIGO}} and {{Virgo}} during the
  {{First}} and {{Second Observing Runs}}.
\newblock {\em Phys. Rev. X}, 9(3):031040, September 2019.

\bibitem{abbottGW170608Observation192017}
B.~P. Abbott et al.
\newblock {{GW170608}}: {{Observation}} of a 19 {{Solar}}-mass {{Binary Black
  Hole Coalescence}}.
\newblock {\em ApJL}, 851(2):L35, December 2017.

\bibitem{ligoscientificandvirgocollaborationGW170104Observation50SolarMass2017}
{LIGO Scientific and Virgo Collaboration} et al.
\newblock {{GW170104}}: {{Observation}} of a 50-{{Solar}}-{{Mass Binary Black
  Hole Coalescence}} at {{Redshift}} 0.2.
\newblock {\em Phys. Rev. Lett.}, 118(22):221101, June 2017.

\bibitem{ligoscientificcollaborationandvirgocollaborationGW151226ObservationGravitational2016}
{LIGO Scientific Collaboration and Virgo Collaboration} et al.
\newblock {{GW151226}}: {{Observation}} of {{Gravitational Waves}} from a
  22-{{Solar}}-{{Mass Binary Black Hole Coalescence}}.
\newblock {\em Phys. Rev. Lett.}, 116(24):241103, June 2016.

\bibitem{ligoscientificcollaborationandvirgocollaborationObservationGravitationalWaves2016}
{LIGO Scientific Collaboration and Virgo Collaboration} et al.
\newblock Observation of {{Gravitational Waves}} from a {{Binary Black Hole
  Merger}}.
\newblock {\em Phys. Rev. Lett.}, 116(6):061102, February 2016.

\bibitem{theligoscientificcollaborationGW190412ObservationBinaryBlackHole2020a}
T.~L.~S. Collaboration et al.
\newblock {{GW190412}}: {{Observation}} of a {{Binary}}-{{Black}}-{{Hole
  Coalescence}} with {{Asymmetric Masses}}.
\newblock {\em arXiv:2004.08342 [astro-ph, physics:gr-qc]}, April 2020.

\bibitem{2017arXiv170200786A}
P.~{Amaro-Seoane} et al.
\newblock {Laser Interferometer Space Antenna}.
\newblock {\em arXiv e-prints}, page arXiv:1702.00786, February 2017.

\bibitem{barackLISACaptureSources2004}
L.~Barack and C.~Cutler.
\newblock {{LISA Capture Sources}}: {{Approximate Waveforms}},
  {{Signal}}-to-{{Noise Ratios}}, and {{Parameter Estimation Accuracy}}.
\newblock {\em Phys. Rev. D}, 69(8):082005, April 2004.

\bibitem{arunMassiveBlackHole2009}
K.~G. Arun et al.
\newblock Massive {{Black Hole Binary Inspirals}}: {{Results}} from the {{LISA
  Parameter Estimation Taskforce}}.
\newblock {\em Class. Quantum Grav.}, 26(9):094027, May 2009.

\bibitem{barausseCanEnvironmentalEffects2014}
E.~Barausse, V.~Cardoso, and P.~Pani.
\newblock Can environmental effects spoil precision gravitational-wave
  astrophysics?
\newblock {\em Phys. Rev. D}, 89(10):104059, May 2014.

\bibitem{gairConstrainingPropertiesBlack2011}
J.~R. Gair, A.~Sesana, E.~Berti, and M.~Volonteri.
\newblock Constraining properties of the black hole population using {{LISA}}.
\newblock {\em Class. Quantum Grav.}, 28(9):094018, May 2011.

\bibitem{gairLISAExtrememassratioInspiral2010}
J.~R. Gair, C.~Tang, and M.~Volonteri.
\newblock {{LISA}} extreme-mass-ratio inspiral events as probes of the black
  hole mass function.
\newblock {\em Phys. Rev. D}, 81(10):104014, May 2010.

\bibitem{amaro-seoaneIntermediateExtremeMassRatio2007}
P.~{Amaro-Seoane} et al.
\newblock Intermediate and {{Extreme Mass}}-{{Ratio Inspirals}} --
  {{Astrophysics}}, {{Science Applications}} and {{Detection}} using {{LISA}}.
\newblock {\em Class. Quantum Grav.}, 24(17):R113--R169, September 2007.

\bibitem{gairTestingGeneralRelativity2013}
J.~R. Gair, M.~Vallisneri, S.~L. Larson, and J.~G. Baker.
\newblock Testing {{General Relativity}} with {{Low}}-{{Frequency}},
  {{Space}}-{{Based Gravitational}}-{{Wave Detectors}}.
\newblock {\em Living Rev. Relativ.}, 16(1):7, December 2013.

\bibitem{barackUsingLISAEMRI2007}
L.~Barack and C.~Cutler.
\newblock Using {{LISA EMRI}} sources to test off-{{Kerr}} deviations in the
  geometry of massive black holes.
\newblock {\em Phys. Rev. D}, 75(4):042003, February 2007.

\bibitem{macleodPrecisionHubbleConstant2008}
C.~L. MacLeod and C.~J. Hogan.
\newblock Precision of {{Hubble}} constant derived using black hole binary
  absolute distances and statistical redshift information.
\newblock {\em Phys. Rev. D}, 77(4):043512, February 2008.

\bibitem{laghi2021gravitational}
D.~Laghi et al.
\newblock Gravitational wave cosmology with extreme mass-ratio inspirals, 2021.

\bibitem{gairEventRateEstimates2004}
J.~R. Gair et al.
\newblock Event rate estimates for {{LISA}} extreme mass ratio capture sources.
\newblock {\em Class. Quantum Grav.}, 21(20):S1595--S1606, October 2004.

\bibitem{gairProbingBlackHoles2009}
J.~R. Gair.
\newblock Probing black holes at low redshift using {{LISA EMRI}} observations.
\newblock {\em Class. Quantum Grav.}, 26(9):094034, April 2009.

\bibitem{babakScienceSpacebasedInterferometer2017}
S.~Babak et al.
\newblock Science with the space-based interferometer {{LISA}}. {{V}}:
  {{Extreme}} mass-ratio inspirals.
\newblock {\em Phys. Rev. D}, 95(10):103012, May 2017.

\bibitem{pan2021formation}
Z.~Pan and H.~Yang.
\newblock Formation rate of extreme mass ratio inspirals in active galactic
  nucleus, 2021.

\bibitem{amaro-seoaneRelativisticDynamicsExtreme2018}
P.~{Amaro-Seoane}.
\newblock Relativistic {{Dynamics}} and {{Extreme Mass Ratio Inspirals}}.
\newblock {\em Living Rev Relativ}, 21(1):4, December 2018.

\bibitem{babakMockLISAData2008a}
S.~Babak et al.
\newblock The {{Mock LISA Data Challenges}}: From {{Challenge 1B}} to
  {{Challenge}} 3.
\newblock {\em Class. Quantum Grav.}, 25(18):184026, September 2008.

\bibitem{babakAlgorithmDetectionExtreme2009}
S.~Babak, J.~R. Gair, and E.~K. Porter.
\newblock An algorithm for the detection of extreme mass ratio inspirals in
  {{LISA}} data.
\newblock {\em Class. Quantum Grav.}, 26(13):135004, June 2009.

\bibitem{babakMockLISAData2010}
S.~Babak et al.
\newblock The {{Mock LISA Data Challenges}}: From challenge 3 to challenge 4.
\newblock {\em Class. Quantum Grav.}, 27(8):084009, April 2010.

\bibitem{Chua:2020stf}
A.~J. Chua, M.~L. Katz, N.~Warburton, and S.~A. Hughes.
\newblock {Rapid generation of fully relativistic extreme-mass-ratio-inspiral
  waveform templates for LISA data analysis}.
\newblock 8 2020.

\bibitem{Katz:2021yft}
M.~L. Katz et al.
\newblock {FastEMRIWaveforms: New tools for millihertz gravitational-wave data
  analysis}.
\newblock 4 2021.

\bibitem{PhysRevD.55.3457}
Y.~Mino, M.~Sasaki, and T.~Tanaka.
\newblock Gravitational radiation reaction to a particle motion.
\newblock {\em Phys. Rev. D}, 55:3457--3476, Mar 1997.

\bibitem{PhysRevD.56.3381}
T.~C. Quinn and R.~M. Wald.
\newblock Axiomatic approach to electromagnetic and gravitational radiation
  reaction of particles in curved spacetime.
\newblock {\em Phys. Rev. D}, 56:3381--3394, Sep 1997.

\bibitem{flanaganTransientResonancesInspirals2012}
E.~E. Flanagan and T.~Hinderer.
\newblock Transient resonances in the inspirals of point particles into black
  holes.
\newblock {\em Phys. Rev. Lett.}, 109(7):071102, August 2012.

\bibitem{berryImportanceTransientResonances2016}
C.~P.~L. Berry, R.~H. Cole, P.~Ca{\~n}izares, and J.~R. Gair.
\newblock Importance of transient resonances in extreme-mass-ratio inspirals.
\newblock {\em Phys. Rev. D}, 94(12):124042, December 2016.

\bibitem{vandemeentConditionsSustainedOrbital2014}
M.~{van de Meent}.
\newblock Conditions for {{Sustained Orbital Resonances}} in {{Extreme Mass
  Ratio Inspirals}}.
\newblock {\em Phys. Rev. D}, 89(8):084033, April 2014.

\bibitem{vandemeentResonantlyEnhancedKicks2014}
M.~{van de Meent}.
\newblock Resonantly enhanced kicks from equatorial small mass-ratio inspirals.
\newblock {\em Phys. Rev. D}, 90(4):044027, August 2014.

\bibitem{bongaTidalResonanceExtreme2019}
B.~Bonga, H.~Yang, and S.~A. Hughes.
\newblock Tidal resonance in extreme mass-ratio inspirals.
\newblock {\em Phys. Rev. Lett.}, 123(10):101103, September 2019.

\bibitem{Brink_2015}
J.~Brink, M.~Geyer, and T.~Hinderer.
\newblock Orbital resonances around black holes.
\newblock {\em Physical Review Letters}, 114(8), Feb 2015.

\bibitem{lukesgerakopoulos2021nonlinear}
G.~Lukes-Gerakopoulos and V.~Witzany.
\newblock Non-linear effects in emri dynamics and their imprints on
  gravitational waves, 2021.

\bibitem{flanaganResonantlyEnhancedDiminished2014}
E.~E. Flanagan, S.~A. Hughes, and U.~Ruangsri.
\newblock Resonantly enhanced and diminished strong-field gravitational-wave
  fluxes.
\newblock {\em Phys. Rev. D}, 89(8):084028, April 2014.

\bibitem{ruangsriCensusTransientOrbital2014}
U.~Ruangsri and S.~A. Hughes.
\newblock A census of transient orbital resonances encountered during binary
  inspiral.
\newblock {\em Phys. Rev. D}, 89(8):084036, April 2014.

\bibitem{PhysRevD.88.023002}
R.~Grossman, J.~Levin, and G.~Perez-Giz.
\newblock Faster computation of adiabatic extreme mass-ratio inspirals using
  resonances.
\newblock {\em Phys. Rev. D}, 88:023002, Jul 2013.

\bibitem{PhysRevD.83.104024}
C.~M. Hirata.
\newblock Resonant recoil in extreme mass ratio binary black hole mergers.
\newblock {\em Phys. Rev. D}, 83:104024, May 2011.

\bibitem{10.1093/ptep/pty136}
S.~Isoyama et al.
\newblock {Flux-balance formulae for extreme mass-ratio inspirals}.
\newblock {\em Progress of Theoretical and Experimental Physics}, 2019(1), 01
  2019.
\newblock 013E01.

\bibitem{babakKludgeGravitationalWaveforms2008}
S.~Babak et al.
\newblock ``kludge'' gravitational waveforms for a test-body orbiting a kerr
  black hole.
\newblock {\em Phys. Rev. D}, 75:024005, Jan 2007.

\bibitem{gairImprovedApproximateInspirals2006}
J.~R. Gair and K.~Glampedakis.
\newblock Improved approximate inspirals of test-bodies into {{Kerr}} black
  holes.
\newblock {\em Phys. Rev. D}, 73(6):064037, March 2006.

\bibitem{carterGlobalStructureKerr1968}
B.~Carter.
\newblock Global {{Structure}} of the {{Kerr Family}} of {{Gravitational
  Fields}}.
\newblock {\em Phys. Rev.}, 174(5):1559--1571, October 1968.

\bibitem{PhysRevD.67.084027}
Y.~Mino.
\newblock Perturbative approach to an orbital evolution around a supermassive
  black hole.
\newblock {\em Phys. Rev. D}, 67:084027, Apr 2003.

\bibitem{barackSelfforceRadiationReaction2019}
L.~Barack and A.~Pound.
\newblock Self-force and radiation reaction in general relativity.
\newblock {\em Rep. Prog. Phys.}, 82(1):016904, January 2019.

\bibitem{hindererTwoTimescaleAnalysis2008}
T.~Hinderer and E.~E. Flanagan.
\newblock Two timescale analysis of extreme mass ratio inspirals in {{Kerr}}.
  {{I}}. {{Orbital Motion}}.
\newblock {\em Phys. Rev. D}, 78(6):064028, September 2008.

\bibitem{fioraniLiouvilleArnoldNekhoroshev2003}
E.~Fiorani, G.~Giachetta, and G.~Sardanashvily.
\newblock The {{Liouville Arnold Nekhoroshev}} theorem for non-compact
  invariant manifolds.
\newblock {\em J. Phys. A: Math. Gen.}, 36(7):L101--L107, February 2003.

\bibitem{schmidtCelestialMechanicsKerr2002}
W.~Schmidt.
\newblock Celestial mechanics in {{Kerr}} spacetime.
\newblock {\em Class. Quantum Grav.}, 19(10):2743--2764, May 2002.

\bibitem{Fujita_2009}
R.~Fujita and W.~Hikida.
\newblock Analytical solutions of bound timelike geodesic orbits in kerr
  spacetime.
\newblock {\em Classical and Quantum Gravity}, 26(13):135002, Jun 2009.

\bibitem{chuaFastFiducialAugmented2017}
A.~J.~K. Chua, C.~J. Moore, and J.~R. Gair.
\newblock The {{Fast}} and the {{Fiducial}}: {{Augmented}} kludge waveforms for
  detecting extreme-mass-ratio inspirals.
\newblock {\em Phys. Rev. D}, 96(4):044005, August 2017.

\bibitem{vandemeentFastSelfforcedInspirals2018}
M.~{van de Meent} and N.~Warburton.
\newblock Fast {{Self}}-forced {{Inspirals}}.
\newblock {\em Class. Quantum Grav.}, 35(14):144003, July 2018.

\bibitem{PhysRevD.69.044015}
S.~Drasco and S.~A. Hughes.
\newblock Rotating black hole orbit functionals in the frequency domain.
\newblock {\em Phys. Rev. D}, 69:044015, Feb 2004.

\bibitem{poundLimitationsAdiabaticApproximation2005}
A.~Pound, E.~Poisson, and B.~G. Nickel.
\newblock Limitations of the adiabatic approximation to the gravitational
  self-force.
\newblock {\em Phys. Rev. D}, 72(12):124001, December 2005.

\bibitem{cutlerAngularResolutionLISA1998}
C.~Cutler.
\newblock Angular {{Resolution}} of the {{LISA Gravitational Wave Detector}}.
\newblock {\em Phys. Rev. D}, 57(12):7089--7102, June 1998.

\bibitem{whittleAnalysisMultipleStationary1953}
P.~Whittle.
\newblock The {{Analysis}} of {{Multiple Stationary Time Series}}.
\newblock {\em Journal of the Royal Statistical Society. Series B
  (Methodological)}, 15(1):125--139, 1953.

\bibitem{robsonConstructionUseLISA2019}
T.~Robson, N.~Cornish, and C.~Liu.
\newblock The construction and use of {{LISA}} sensitivity curves.
\newblock {\em Class. Quantum Grav.}, 36(10):105011, May 2019.

\bibitem{vallisneriStealthBiasGravitationalWave2013}
M.~Vallisneri and N.~Yunes.
\newblock Stealth {{Bias}} in {{Gravitational}}-{{Wave Parameter Estimation}}.
\newblock {\em Phys. Rev. D}, 87(10):102002, May 2013.

\bibitem{mihaylovTransitionEMRIsResonance2017}
D.~P. Mihaylov and J.~Gair.
\newblock Transition of {{EMRIs}} through resonance: Corrections to higher
  order in the on-resonance flux modification.
\newblock {\em Journal of Mathematical Physics}, 58(11):112501, November 2017.

\bibitem{1455106}
F.~J. {Harris}.
\newblock On the use of windows for harmonic analysis with the discrete fourier
  transform.
\newblock {\em Proceedings of the IEEE}, 66(1):51--83, 1978.

\bibitem{6768513}
R.~B. {Blackman} and J.~W. {Tukey}.
\newblock The measurement of power spectra from the point of view of
  communications engineering — part i.
\newblock {\em The Bell System Technical Journal}, 37(1):185--282, 1958.

\bibitem{SASPWEB2011}
J.~O. Smith.
\newblock {\em Spectral Audio Signal Processing}.
\newblock
  \htmladdnormallink{\texttt{http:}}{http://ccrma.stanford.edu/~jos/sasp/}\texttt{//\-ccrma.stanford.edu/\-\~{}jos/\-sasp/}.
\newblock online book, 2011 edition.

\bibitem{BlackHolePerturbation}
{Black Hole Perturbation Toolkit}.
\newblock (\href{http://bhptoolkit.org/}{bhptoolkit.org}).

\bibitem{mcclintockSpinNearExtremeKerr2006}
J.~E. McClintock et al.
\newblock The {{Spin}} of the {{Near}}-{{Extreme Kerr Black Hole GRS}}
  1915+105.
\newblock {\em ApJ}, 652(1):518, November 2006.

\bibitem{risalitiRapidlySpinningSupermassive2013}
G.~Risaliti et al.
\newblock A rapidly spinning supermassive black hole at the centre of {{NGC}}
  1365.
\newblock {\em Nature}, 494(7438):449--451, February 2013.

\bibitem{gouEXTREMESPINBLACK2011}
L.~Gou et al.
\newblock {{THE EXTREME SPIN OF THE BLACK HOLE IN CYGNUS X}}-1.
\newblock {\em ApJ}, 742(2):85, November 2011.

\bibitem{Flanagan:1997kp}
E.~E. Flanagan and S.~A. Hughes.
\newblock {Measuring gravitational waves from binary black hole coalescences:
  2. The Waves' information and its extraction, with and without templates}.
\newblock {\em Phys. Rev. D}, 57:4566--4587, 1998.

\bibitem{PhysRevD.71.104016}
M.~Miller.
\newblock Accuracy requirements for the calculation of gravitational waveforms
  from coalescing compact binaries in numerical relativity.
\newblock {\em Phys. Rev. D}, 71:104016, May 2005.

\bibitem{Lindblom:2008cm}
L.~Lindblom, B.~J. Owen, and D.~A. Brown.
\newblock {Model Waveform Accuracy Standards for Gravitational Wave Data
  Analysis}.
\newblock {\em Phys. Rev. D}, 78:124020, 2008.

\bibitem{2015PhRvD..91l4062G}
J.~R. {Gair} and C.~J. {Moore}.
\newblock {Quantifying and mitigating bias in inference on gravitational wave
  source populations}.
\newblock {\em \prd}, 91(12):124062, June 2015.

\bibitem{2020SciPy-NMeth}
P.~Virtanen et al.
\newblock {{SciPy} 1.0: Fundamental Algorithms for Scientific Computing in
  Python}.
\newblock {\em Nature Methods}, 17:261--272, 2020.

\bibitem{vallisneriUseAbuseFisher2008}
M.~Vallisneri.
\newblock Use and {{Abuse}} of the {{Fisher Information Matrix}} in the
  {{Assessment}} of {{Gravitational}}-{{Wave Parameter}}-{{Estimation
  Prospects}}.
\newblock {\em Phys. Rev. D}, 77(4):042001, February 2008.

\bibitem{Burke_2020}
O.~Burke, J.~R. Gair, J.~Simón, and M.~C. Edwards.
\newblock Constraining the spin parameter of near-extremal black holes using
  lisa.
\newblock {\em Physical Review D}, 102(12), Dec 2020.

\end{thebibliography}

\end{document}